\pdfoutput=1
\documentclass[twocolumn, tighten]{aastex61}
\usepackage{amsmath}

\begin{document}
\shorttitle{Sonneberg plate photometry for Boyajian's Star in two passbands}
\shortauthors{Hippke et al.}
\title{Sonneberg plate photometry for Boyajian's Star in two passbands}

\author{Michael Hippke}
\affiliation{Luiter Stra{\ss}e 21b, 47506 Neukirchen-Vluyn, Germany}

\author{Peter Kroll}
\affiliation{Sonneberg Observatory, Sternwartestr. 32, 96515 Sonneberg, Germany}

\author{Frank Matthai}
\affiliation{Sonneberg Observatory, Sternwartestr. 32, 96515 Sonneberg, Germany}

\author{Daniel Angerhausen}
\affiliation{NASA Postdoctoral Program Fellow, NASA Goddard Space Flight Center, Exoplanets \& Stellar Astrophysics Laboratory, Code 667, Greenbelt, MD 20771}

\author{Taavi Tuvikene}
\affiliation{Tartu Observatory, Observatooriumi 1, 61602 T\~oravere, Estonia}

\author{Keivan G. Stassun}
\affiliation{Department of Physics and Astronomy, Vanderbilt University, Nashville, TN 37235, USA}

\author{Elena Roshchina}
\affiliation{Main (Pulkovo) Astronomical Observatory of RAS, 196140, Russia, Saint-Petersburg, Pulkovskoye shosse 65}

\author{Tatyana Vasileva}
\affiliation{Main (Pulkovo) Astronomical Observatory of RAS, 196140, Russia, Saint-Petersburg, Pulkovskoye shosse 65}

\author{Igor Izmailov}
\affiliation{Main (Pulkovo) Astronomical Observatory of RAS, 196140, Russia, Saint-Petersburg, Pulkovskoye shosse 65}

\author{Nikolay N. Samus}
\affiliation{Institute of Astronomy, Russian Academy of Sciences, 48, Pyatnitskaya Str., Moscow 119017, Russia}
\affiliation{Sternberg Astronomical Institute, Lomonosov Moscow University, 13, University Ave., Moscow 119234, Russia}

\author{Elena N. Pastukhova}
\affiliation{Institute of Astronomy, Russian Academy of Sciences, 48, Pyatnitskaya Str., Moscow 119017, Russia}

\author{Ivan Bryukhanov}
\affiliation{Group Betelgeuse, Republic Center of Technical Creativity of Pupils, 12 Makaionak Street, Minsk, 220023, Belarus}

\author{Michael B. Lund}
\affiliation{Department of Physics and Astronomy, Vanderbilt University, Nashville, TN 37235, USA}

\begin{abstract}
The F3 main sequence star KIC 8462852 (Boyajian's Star) showed deep (up to 20\%) day-long brightness dips of unknown cause during the 4 years of the \textit{Kepler} mission. A 0.164~mag (16\%) dimming between 1890 and 1990 was claimed, based on the analysis of photographic plates from the Harvard Observatory. We have gathered an independent set of historic plates from Sonneberg Observatory, Germany, covering the years 1934 -- 1995. With 861 magnitudes in B, and 397 magnitudes in V, we find the star to be of constant brightness within 0.03~mag per century (3\%). Consistent outcomes are found using by-eye estimates of the best 119 plates. Results are supported by data from Sternberg Observatory, Moscow, which show the star as constant between 1895 and 1995. The previously claimed century-long dimming is inconsistent with our results at the $5\sigma$-level, however the recently reported modest dimming of 3\% in the Kepler data is not inconsistent with our data. We find no periodicities or shorter trends within our limits of 5\% per 5-year bin, but note a possible dimming event on 24 Oct 1978.
\end{abstract}

\section{Introduction}
The \textit{Kepler} Space Telescope's exquisite photo\-metry has allowed for the detection of more than a thousand exoplanets \citep{2010Sci...327..977B}. The data have also been used for a wide range of stellar studies \citep[e.g.,][]{2011ApJ...743..143H,2015ApJ...798...42H}. One very peculiar object is Boyajian's Star (KIC 8462852, TYC 3162-665-1). This is an F3 main-sequence star, which was observed by the NASA Kepler mission from April 2009 to May 2013. An analysis by \citet{2016MNRAS.457.3988B} shows inexplicable series of day-long brightness dips up to 20\%. The behavior has been theorized to originate from a family of large comets \citep{2016ApJ...819L..34B}, or signs of a Dyson sphere \citep{2016ApJ...816...17W}. Subsequent analysis found no narrow-band radio signals \citep{2016ApJ...825..155H} and no periodic pulsed optical signals \citep{2016ApJ...825L...5S,  2016ApJ...818L..33A}. The infrared flux is equally unremarkable \citep{2015ApJ...815L..27L, 2015ApJ...814L..15M,  2016MNRAS.458L..39T}. Recently, the star has been claimed to dim by $0.164$~mag ($\sim16\%$) between 1890 and 1990 \citep{2016ApJ...822L..34S}, and lost $\sim3\%$ of brightness during the 4.25yrs of Kepler mission \citep{2016ApJ...830L..39M}. The century-long dimming has been challenged by \citet{2016ApJ...825...73H} and \citet{2016arXiv160502760L}, who find systematic noise to be the most likely cause.

Further effort has been put in regular cadence photometry of the star from ground-based facilities. In parallel to observations from AAVSO amateurs, a community-financed initiative has collected $100,000$ USD, and currently spends the money purchasing telescope time\footnote{\url{https://www.kickstarter.com/projects/608159144/the-most-mysterious-star-in-the-galaxy}}.

The distance estimate from Gaia's DR1 parallax ($391.4\substack{+122.1 \\ -75.2}$~pc, \citet{2016arXiv160905492H}) can not constrain the absolute magnitude of the star. To resolve the controversy whether this star has dimmed considerably over 130 years (and perhaps more so earlier), it is crucial to gather more data.

\section{Method}
A major issue for long-term photo\-metry spanning 130 years were changes in emulsion, filters, geographic locations, aging of optics, and the use of 16 different telescopes at Harvard \citep{2016ApJ...825...73H}. Such changes might introduce small offsets in zeropoints, on the level of $<0.1$~mag, and are very hard to determine and remove from time series. Also, observations in Harvard ceased in the 1960s during the so-called ``Menzel gap'', when organizational changes reduced the observatory activities. It has been speculated that Boyajian's Star has undergone a dimming event in this decade \citep{2016ApJ...830L..39M}. Data from other sources could shed more light on the star's brightness behavior.

\begin{figure}
\includegraphics[width=0.5\textwidth]{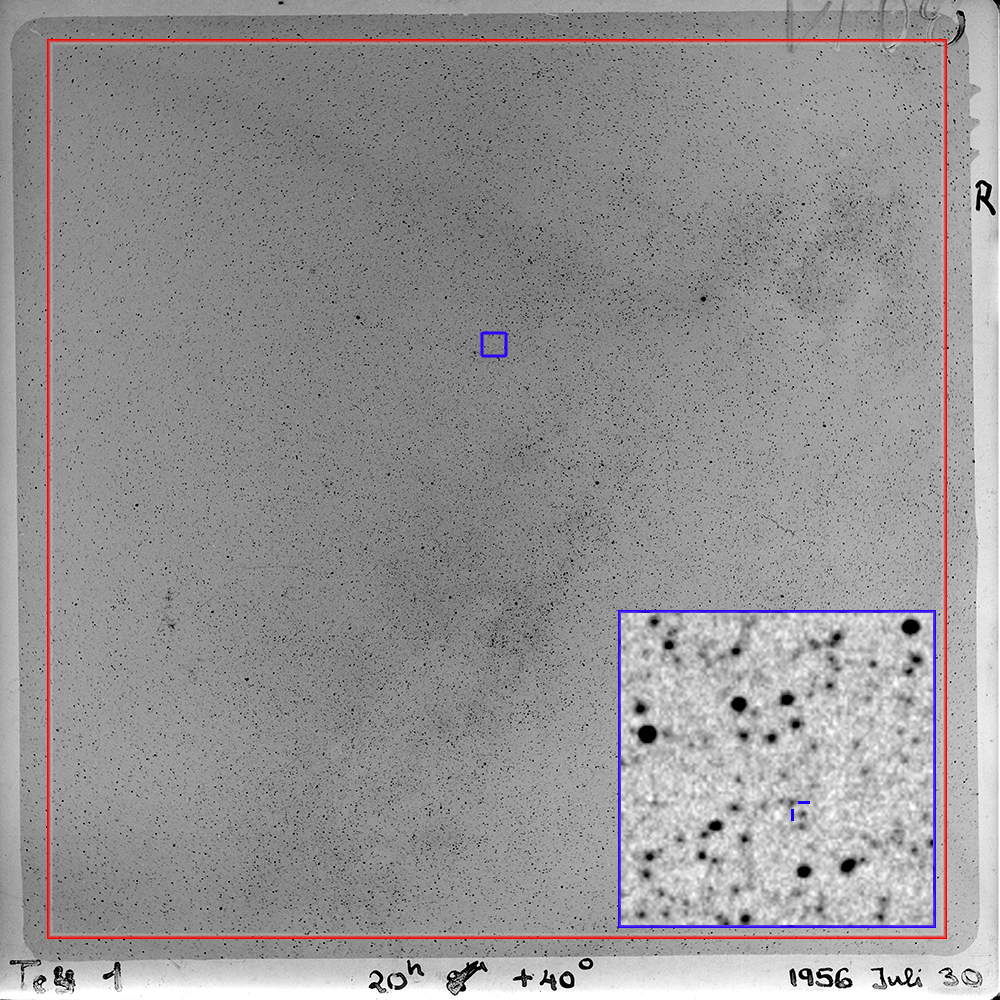}
\caption{\label{fig:raw}A Sonneberg Sky Patrol plate of ave\-rage, representative quality taken on 30 July 1956. Cropped to the inner 90\% of the image (red line) to remove annotations, and some of the vignetting and lens distortion towards the borders. The blue frame inset shows the region around Boyajian's Star (cross-hairs).}
\end{figure}

\begin{figure}
\includegraphics[width=.5\textwidth]{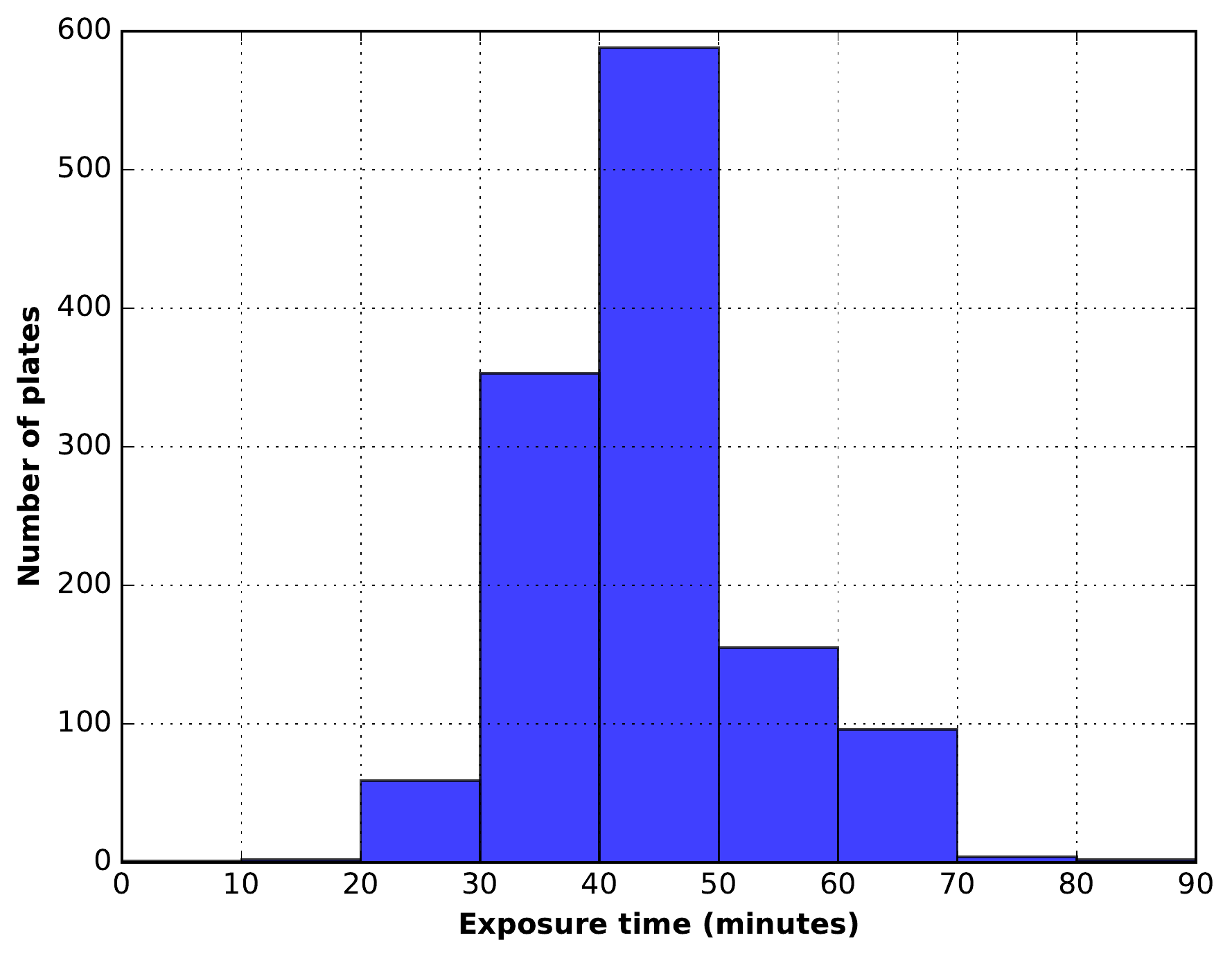}

\includegraphics[width=.5\textwidth]{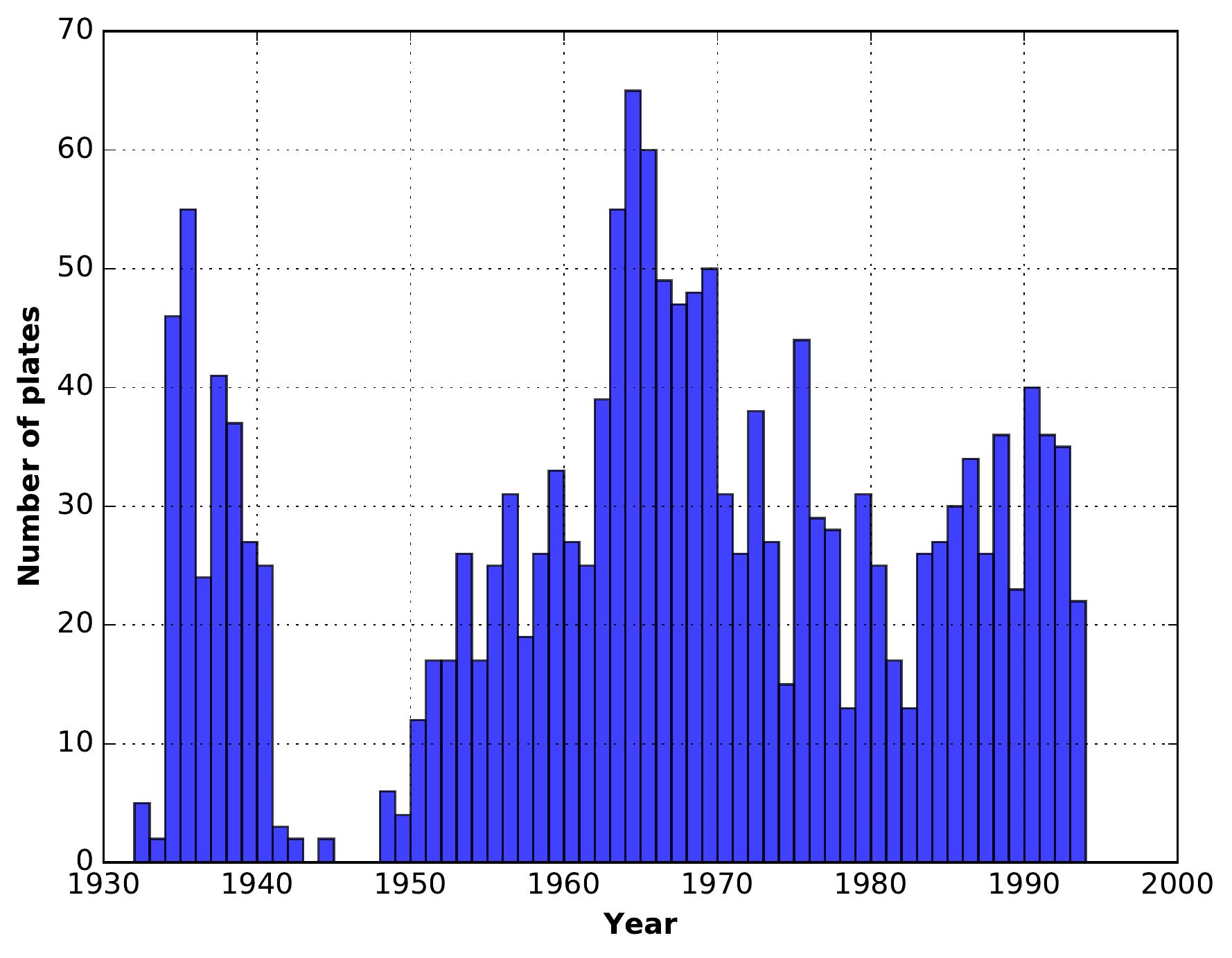}

\includegraphics[width=.5\textwidth]{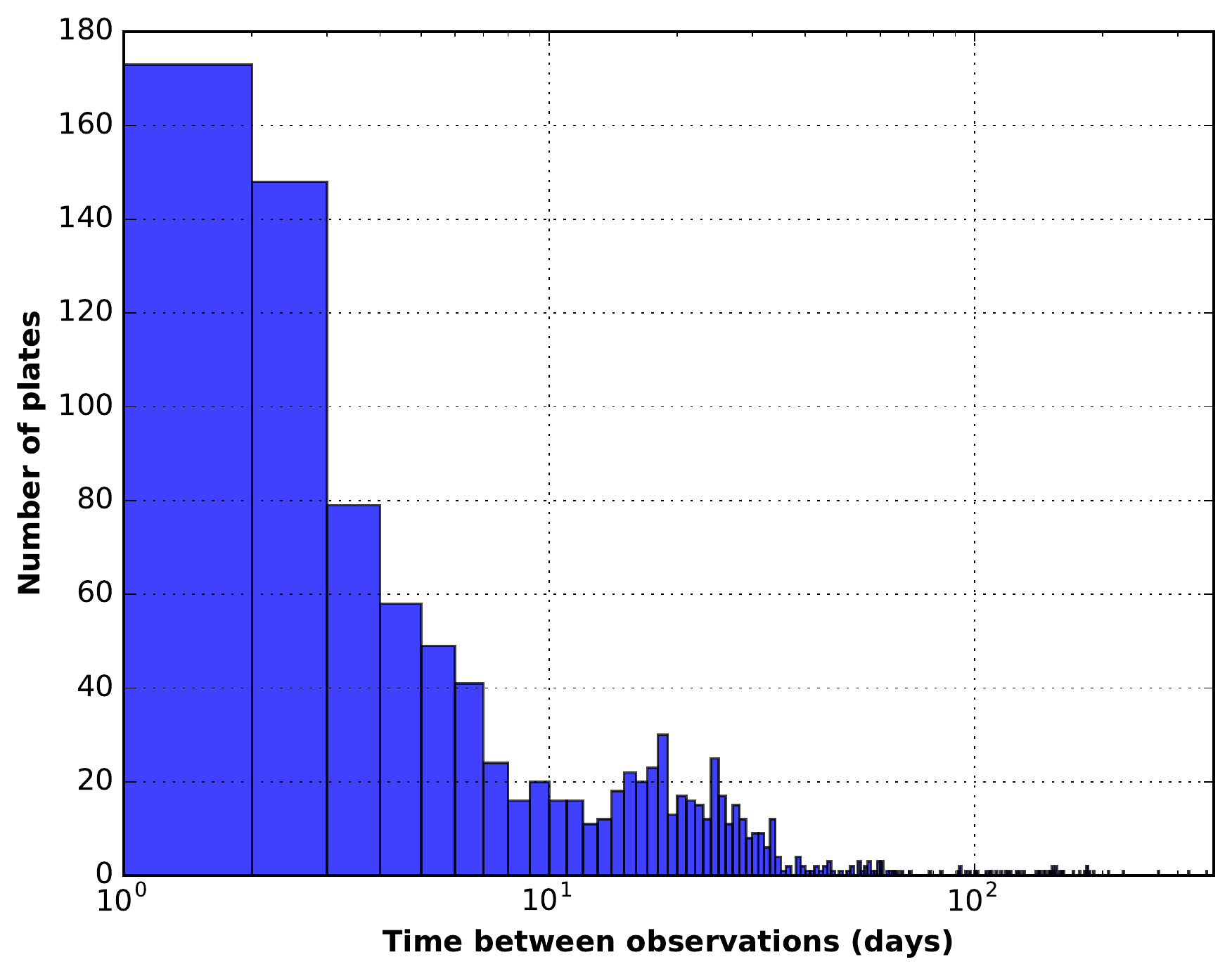}
\caption{\label{fig:hist}Histograms. Top: Most exposure times were between 30 and 60 minutes. Middle: Distribution of plates over the years. The prominent gap in the 1940s is because of the war. Bottom: Time gap between subsequent observations. The majority of plates have a gap of less than a few days, however there are also longer gaps of weeks and months without observations.}
\end{figure}

\begin{figure}
\includegraphics[width=0.5\textwidth]{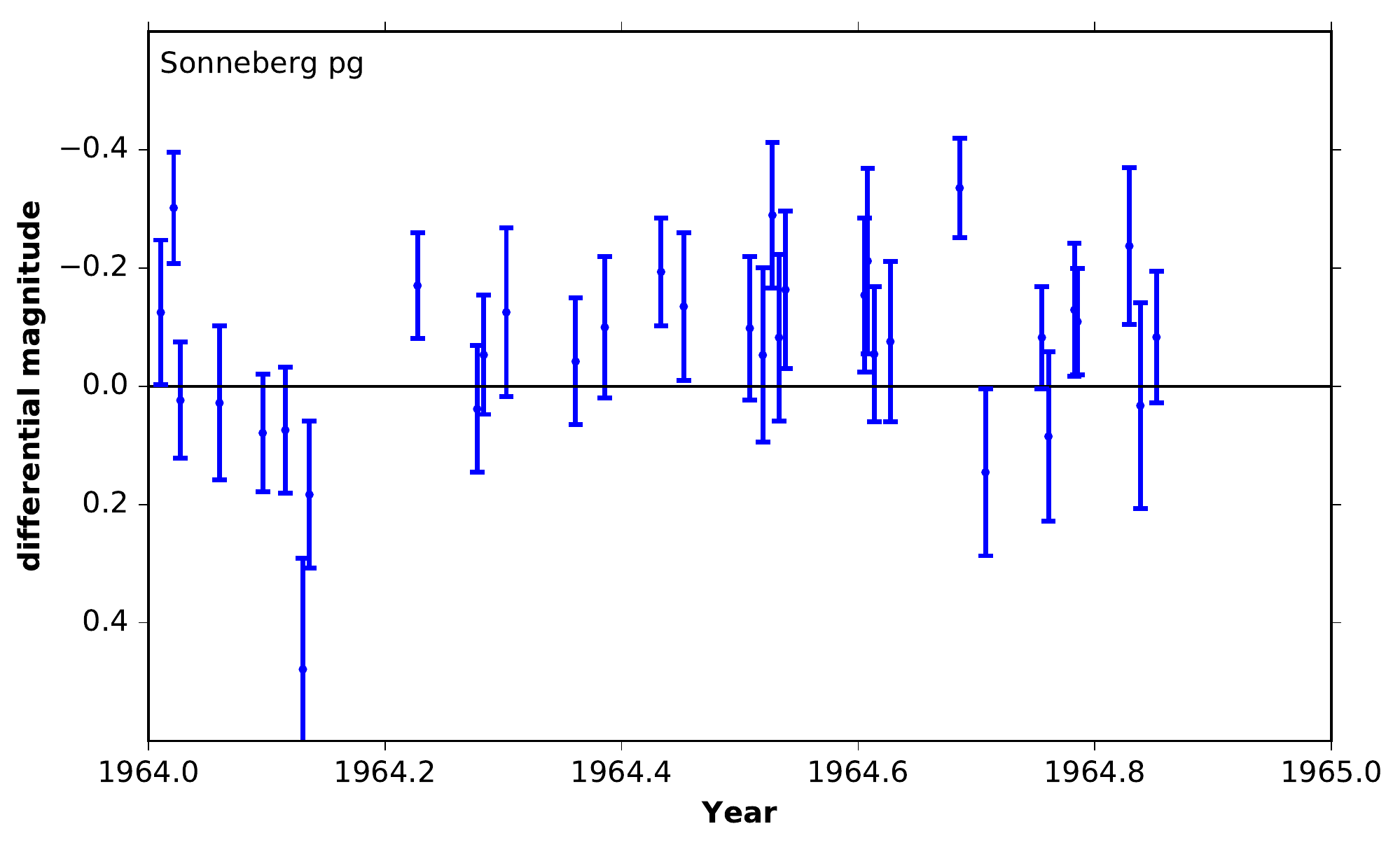}
\caption{\label{fig:oneyear}Data points in \textit{pg} for the year 1964, which has the most plates (total of 65, of which 34 are in \textit{pg}).}
\end{figure}

\subsection{Sonneberg Plate Archive}
\label{sec:archive}
The second largest plate archive in the world, after Harvard, is located at Sonneberg Observatory \citep{1992Stern..68...19B}. The archive was fed by two different observation program: the Sky Patrol, taken with 14 short-focus cameras of the same optics, plate sizes and emulsions, covered the entire sky between about 1935 and 2010 in a very homogeneous manner in two color bands, \textit{pg} (blue) and \textit{pv} (red) \citep{1992Stern..68...19B, 1999AcHA....6...70B}. The deeper Field Patrol aimed at recording about 80 selected areas mostly along the Milky Way and the Galactic North Pole. The archive contains a total of about 275,000 plates.

For this study we have used the Sky Patrol plates of $13\times13$~cm$^2$ size, a scale of 830 arc sec per mm, giving a field size of about $26^{\circ}\times 26^{\circ}$. The limiting magnitudes are of order 14.5~mag in \textit{pg} and 13.5~mag in \textit{pv}. Plate scanning was initially performed with a digitalization machine, but subsequently upgraded with more, and more modern, scanners. Scan resolution was 15 $\mu$m at 16 bit data depth.

We show a plate of average, representative quality in Figure~\ref{fig:raw}. The exposure times were typically 30 to 60 minutes (Figure~\ref{fig:hist}, top panel). The plates were taken between 1934 and 1995, with good sampling between the 1950s and 1990s (Figure~\ref{fig:hist}, middle). The time gap between observations is mostly small (days), but can be weeks or months in some cases (Figure~\ref{fig:hist}, bottom).

\begin{figure*}
\includegraphics[width=.5\textwidth]{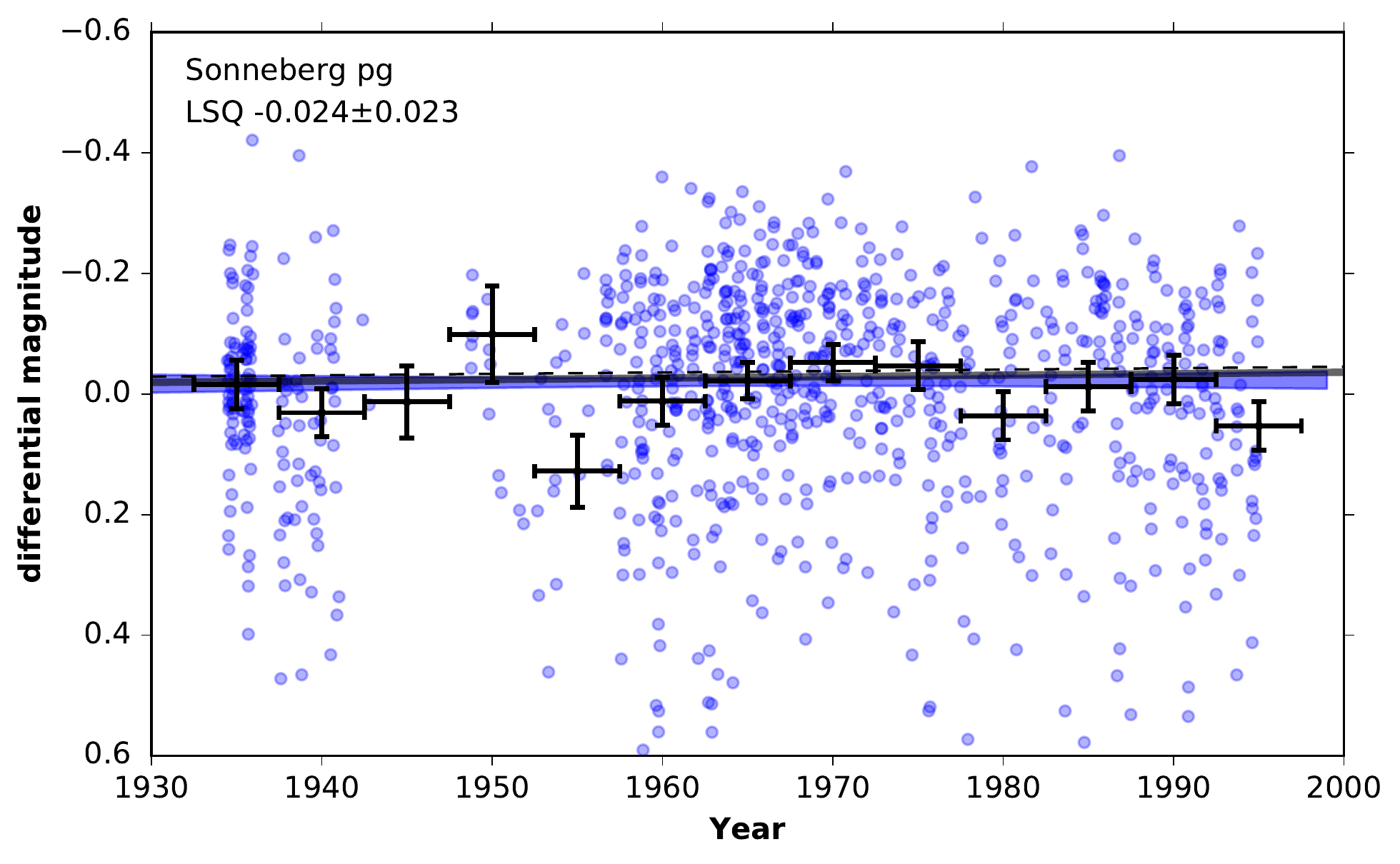}
\includegraphics[width=.5\textwidth]{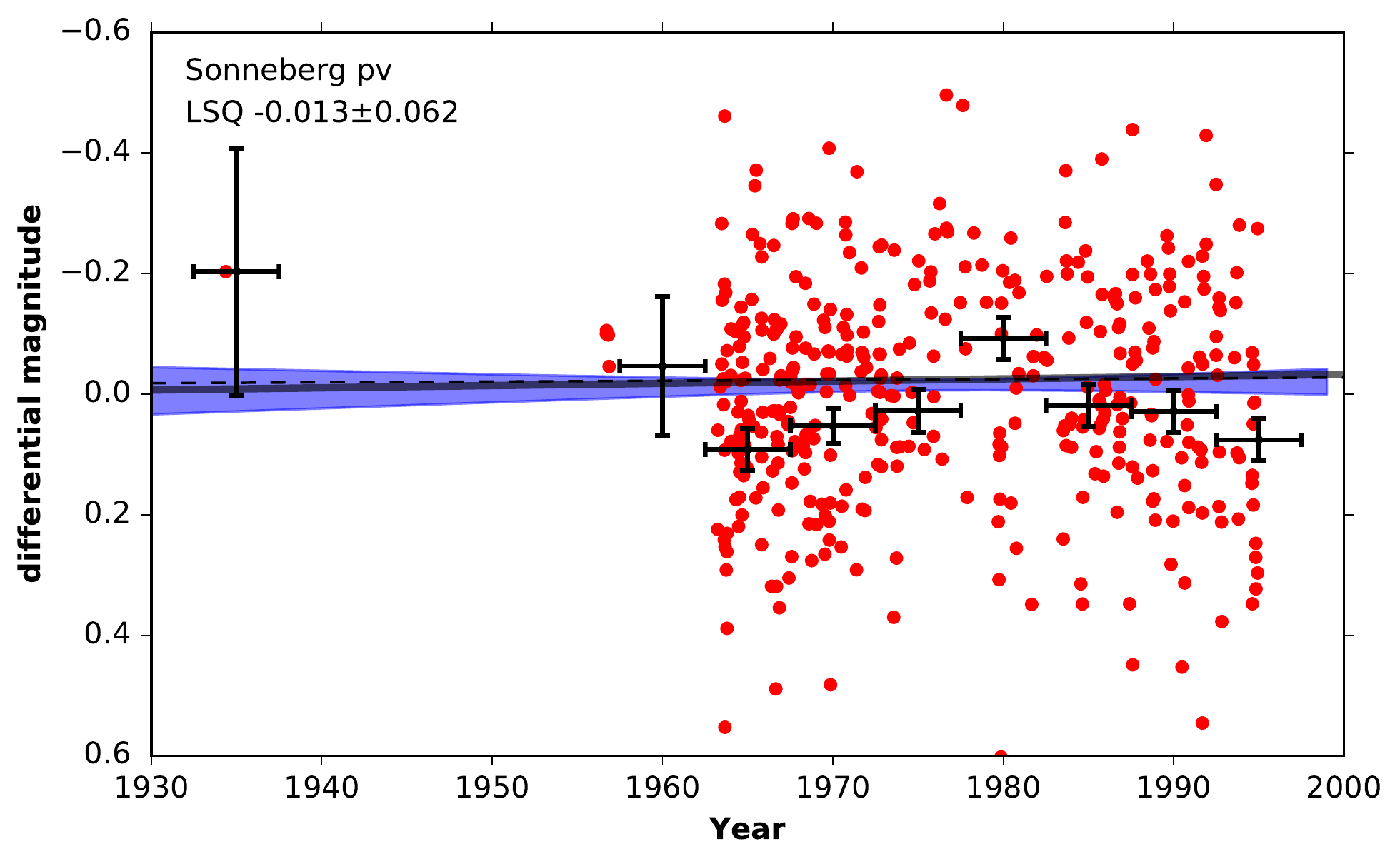}
\caption{\label{fig:sonne_phot}Sonneberg SExtractor photo\-metry for Boyajian's Star in $pg$ (blue) and $pv$ (red) as differential magnitudes with respect to an artificial reference star. Each plot shows a linear least-squares regression (dashed line), a robust regression using the maximum likelihood method (solid gray line), and a linear MCMC regression with $1\sigma$ uncertainties (blue area). The regressions were made using the individual data points including their uncertainties as estimated by SExtractor; the 5-year bins are only shown for visibility.}
\end{figure*}

\begin{figure*}
\includegraphics[width=.5\textwidth]{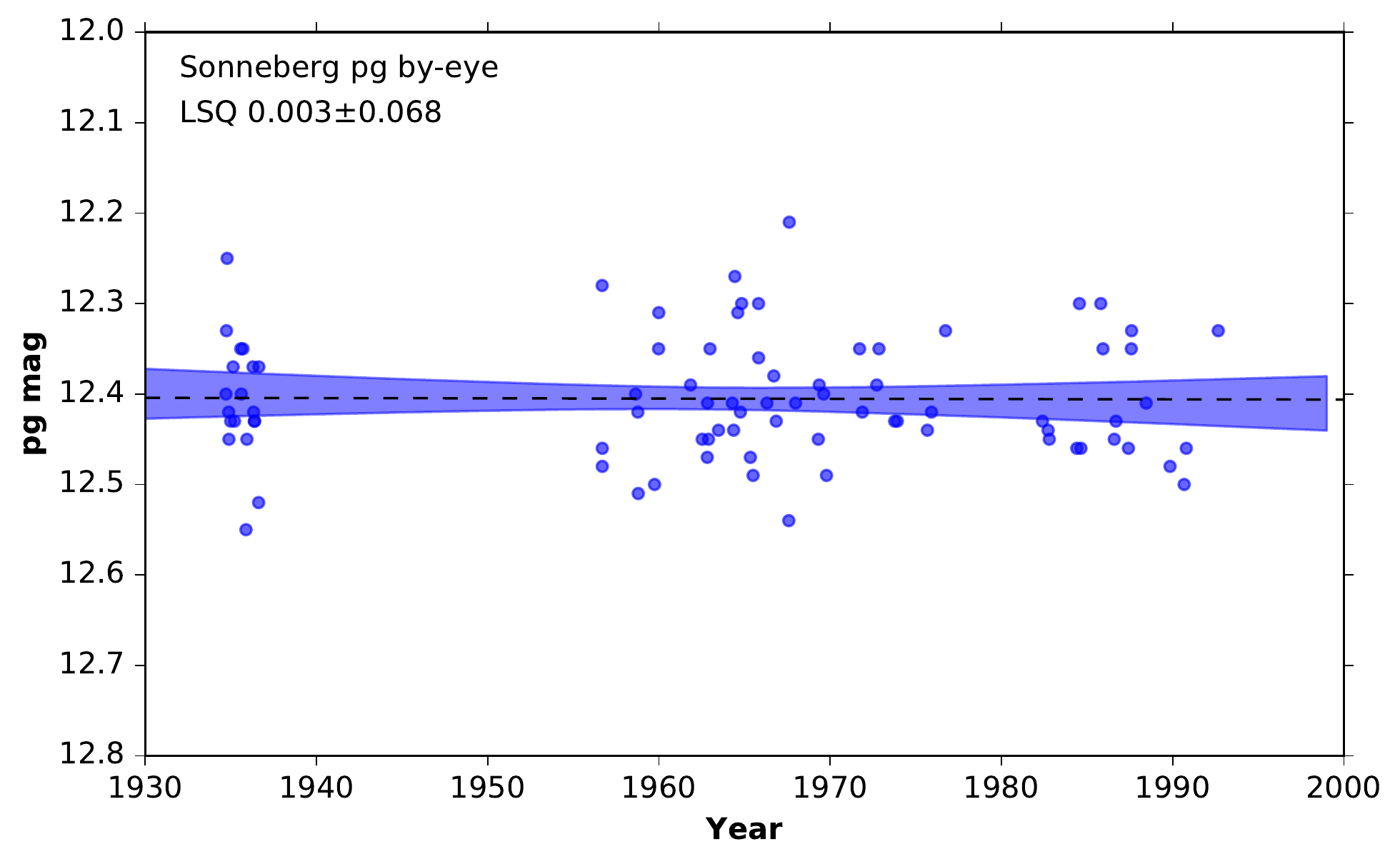}
\includegraphics[width=.5\textwidth]{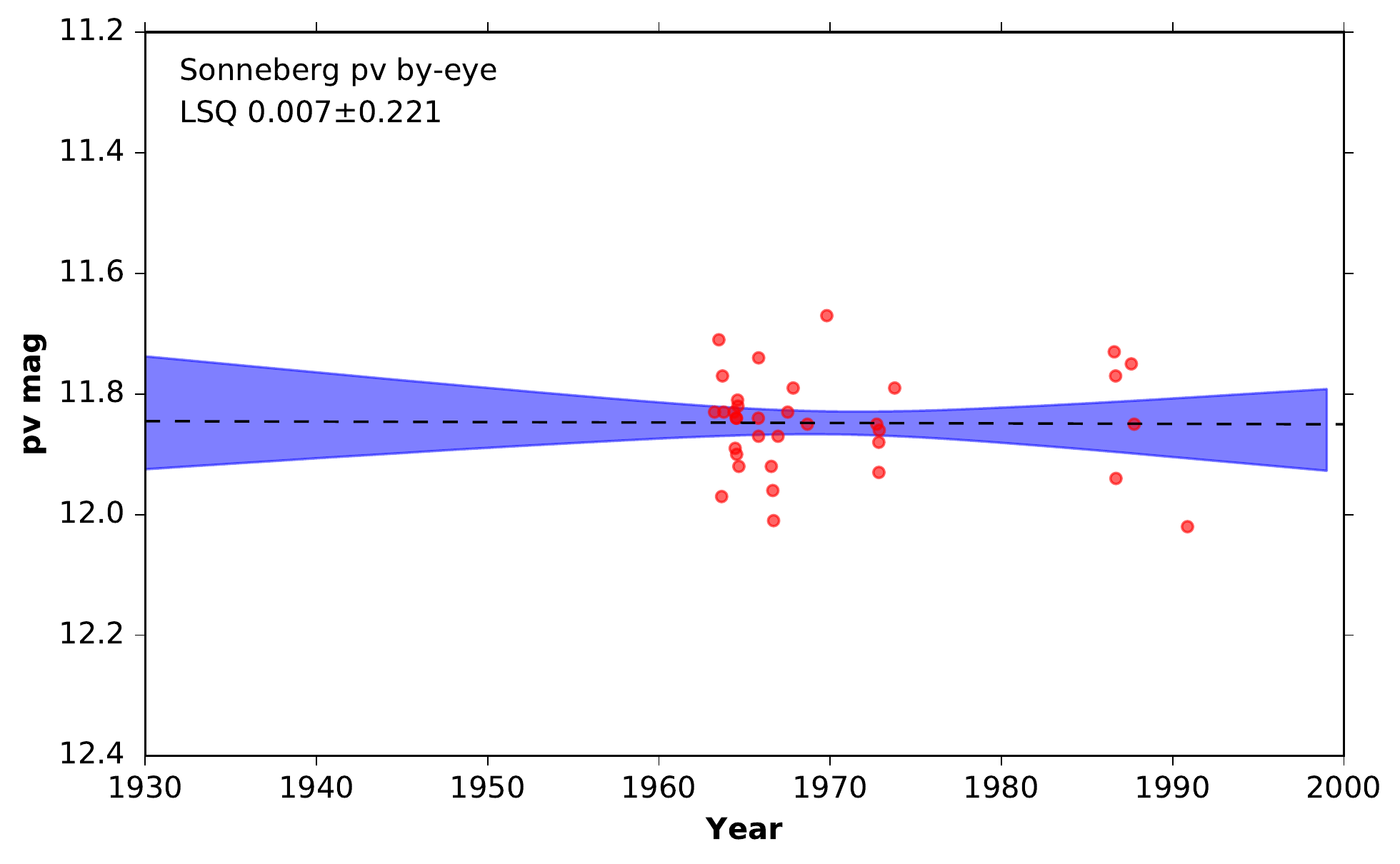}
\caption{\label{fig:byeye}Sonneberg by-eye measurements of the 119 best plates in \textit{pg} (left, 83 plates) and \textit{pv} (right, 36 plates). Scales are identical to Figure~\ref{fig:sonne_phot} for comparison, but in magnitudes.}
\end{figure*}

\subsection{Sonneberg photo\-metry Pipeline}
To maximize robustness, we use an established pipeline to reduce the images, closely resembling the process of the APPLAUSE project in the digitization of the plate archives in Hamburg, Bamberg and Potsdam \citep{2014aspl.conf..127T}. The plate negatives are inverted and cropped to the inner 90\% area of the scanned images to remove the glass borders and annotations. Then, we perform an astrometric solution \citep{2010AJ....139.1782L} using a list of coordinates of the brightest sources as an input and the Tycho-2 catalog as a reference. The resulting astrometric errors are much smaller (few arc sec) than the pixel scale. Boyajian's Star is fortunately located so that its neighbor stars of similar brightness are separated by several pixels (Figure~\ref{fig:raw}), avoiding blends. The star is almost always located within the inner $\sim30$\% from the center of the plate.

After obtaining star coordinates, we perform photo\-metry using the SExtractor program \citep{1996A&AS..117..393B} which identifies all sources above a specified brightness threshold. SExtractor works on scans of photographic plates as well as CCD images; we obtained  the smallest scatter when using a constant circular aperture.

For the calibration, we use nearby comparison stars which are similar in color in order to avoid shifts from emulsion sensitivity changes. We use the Tycho-2 catalog and select stars within 0.5~mag in brightness from our program star, within 1 degree distance, and within 0.1~mag of color difference in Tycho-2 $BT-VT$. This gives a total of 39 comparison stars. In each epoch, we create one artificial ``reference star'' from the mean magnitudes of the comparison stars, and calculate the differential magnitude between the reference star and the program star.

As we are not interested in standard magnitudes, we did not tie the mean magnitudes to a catalogue value. In the future, one might determine the color response of Sonneberg $pg$ and $pv$ and shift the magnitudes slightly for these colors. We estimate the differences between $pg$ and $BT$, and between $pv$ and $BV$, to be $<0.1$~mag. An exemplary part of the resulting light curve for the year 1964 is shown in Figure~\ref{fig:oneyear}.

The SExtractor solution of a single plate contains on ave\-rage 120,000 identified sources, and usable photo\-metry for 100,000 stars. After the calibration, the ave\-rage standard deviation of Boyajian's Star, compared to its mean magnitude, is 0.18~mag per plate in $pg$ and 0.2~mag in $pv$. These values are comparable to other studies using Sonneberg plates (e.g., \citet{2009AJ....138.1846C, 2010PZ.....30....4G,2014ApJ...780L..25J}), and photographic plates in general (e.g., \citet{2010ApJS..187..275S}). The described procedure produces 1263 magnitudes from a total of 1645 plates. As a quality filter, we remove all magnitudes of Boyajian's Star where its calculated astrometric position is located outside of the star image, which is usually $2-3$ pixels. This removes 13 measurements, and eliminates the risk to mix up identifications. The remainder of 369 plates were rejected by SExtractor. Visual examination of 20 of the rejected plates show that glass/emulsion defects and failure of star tracking were the most common issues. We have not attempted to recover these plates.

Visual inspection the raw light curve revealed four $5\sigma$ outliers, resulting from dirt on the glass, which were removed from the final light curve.

For reproducibility and independent re-analysis, we release the relevant raw photo\-metry, metadata, and the code to recreate the Figures as open source\footnote{Data and code are available on Zenodo \dataset[10.5281/zenodo.268011]{https://doi.org/10.5281/zenodo.268011}, \citet{hippke_michael_2017_268011} and Github, \url{http://www.github.com/hippke/sonneberg} (\texttt{commit \#ccf37dc}).}.

\subsection{Visual estimates}
To provide a reference sequence of eye estimates, we selected 119 of the best plates (83 in \textit{pg}, 36 in \textit{pv}) and created cutouts centered on the program star, showing a field of $\sim10\times10$ arc min, similar to \citet[Figure 1]{2016ApJ...822L..34S} with the same annotations using APASS comparison stars in \textit{V} and \textit{B}. The digital images were presented in random order without reference to the observation date. Inspection of the actual glass plates using a microscope would have been preferred, but was not possible because of travel constraints. Magnitudes from the digital images were estimated by leading experts with experience in the estimation of thousands of plates\footnote{Tatyana Vasileva, Nikolay N. Samus, Taavi Tuvikene, Ivan Bryukhanov, Peter Kroll}. Each plate was independently estimated by at least three experts.

\section{Results}
\label{sec:results}
In the following sections, we discuss photo\-metry from Sonneberg as well as other sources. Most magnitudes are reported in $B$, however they have different zeropoints because of the different use of filters and photographic emulsions. Unless otherwise noted, we have not modified the magnitudes, but left them as reported in the given sources. That is, we do not focus on absolute brightness comparisons, but on relative changes for each set of observations individually.

\begin{table*}
\center
\caption{Regression results for the Sonneberg data \label{tab:reg}}
\begin{tabular}{llll}
\tableline
Method & Dataset & $n$ & Slope (mag per century) \\
\tableline
LSQ without uncertainties             & pg & 861 & $-0.007 \pm0.037$ \\
LSQ with uncertainties                & pg & 861 & $-0.024 \pm0.023$ \\
Maximum likelihood with uncertainties & pg & 861 & $-0.024$ (n/a) \\
MCMC with uncertainties               & pg & 861 & $-0.009 \substack{+0.008\\-0.009}$ \\
\tableline
LSQ without uncertainties             & pv & 397 & $-0.002 \pm0.087$ \\
LSQ with uncertainties                & pv & 397 & $-0.013 \pm0.062$ \\
Maximum likelihood with uncertainties & pv & 397 & $-0.037$ (n/a) \\
MCMC with uncertainties               & pv & 397 & $-0.014 \substack{+0.010\\-0.019}$ \\
\tableline
LSQ without uncertainties             & pg + pv & 1258 & $-0.006 \pm0.032$ \\
LSQ with uncertainties                & pg + pv & 1258 & $-0.009 \pm0.020$ \\
Maximum likelihood with uncertainties & pg + pv & 1258 & $-0.017$ (n/a) \\
MCMC with uncertainties               & pg + pv & 1258 & $-0.005 \substack{+0.004\\-0.007}$ \\
\tableline
\end{tabular}
\end{table*}

\subsection{Sonneberg light curve with SExtractor data}
We show the Sonneberg data for Boyajian's Star in Figure~\ref{fig:sonne_phot}. To search for trends, we perform a least-squares linear regression (with, and without the individual uncertainties of the data points as measured by SExtractor). We run regressions separately for each pass-band, and also merged as one dataset.

To account for a non-Gaussian distribution, we perform an additional robust regression using the maximum likelihood method including the individual uncertainties of the data points, and a linear Markov chain Monte Carlo (MCMC) regression using the \textit{emcee} toolkit \citep{2013PASP..125..306F} to estimate the uncertainties.

Results from all methods are consistent within their errors. All methods yield slope parameters consistent with zero (Table~\ref{tab:reg}). This result is consistent with no dimming of the star between 1934 and 1995 and inconsistent with the claimed dimming in \citet{2016ApJ...822L..34S} by $4.9\sigma$ (section~\ref{sec:litcomp}).

\subsection{Sonneberg light curve with by-eye estimates}
Results across experts are very consistent, with individual estimates usually differing 0.1~mag or less. The average scatter of all by-eye estimates is 0.09~mag, less than half compared to all fully automatic SExtractor measurements (0.19~mag), and less than the SExctractor scatter for the selected plates (0.13~mag). We show the averaged result in Figure~\ref{fig:byeye}, which yields $0.003\pm0.068$~mag per century in \textit{pg}, consistent with the SExtractor pipeline, and inconsistent with the claimed dimming. The uncertainty of the \textit{pv} result ($0.007\pm0.221$~mag per century) is too large to be useful, because of the small number of very good plates.

\begin{figure}
\includegraphics[width=0.5\textwidth]{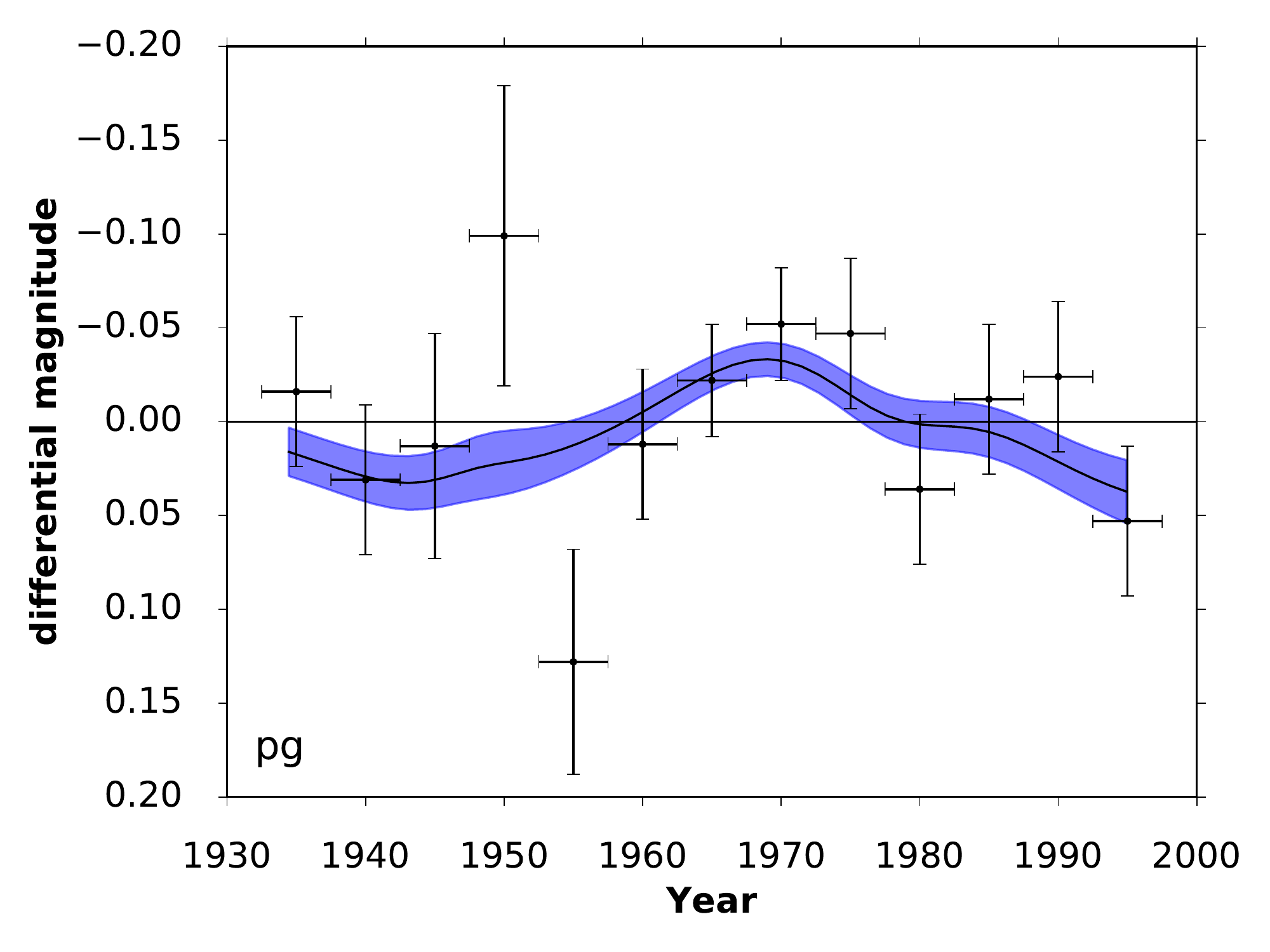}

\includegraphics[width=0.5\textwidth]{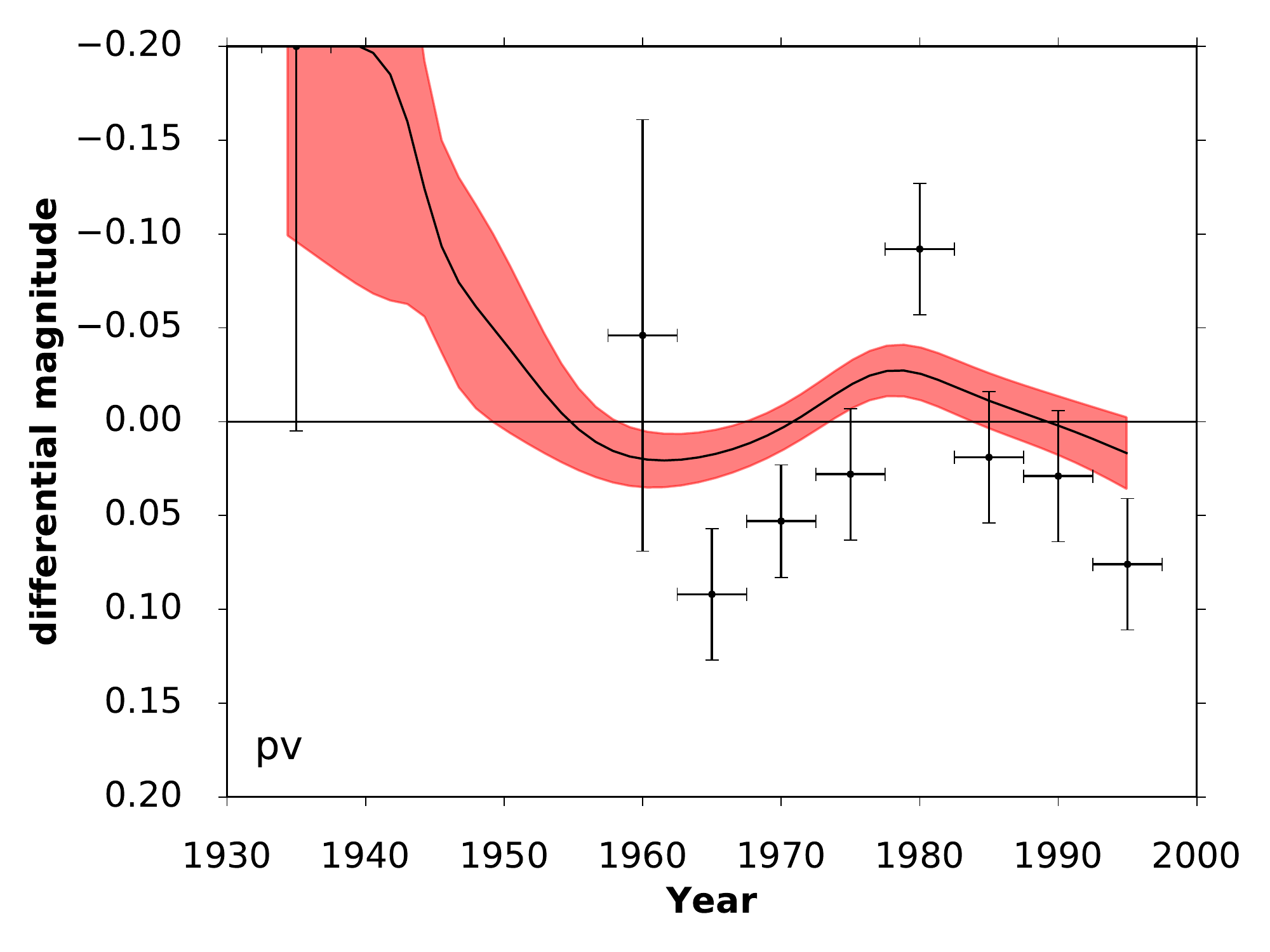}

\includegraphics[width=0.5\textwidth]{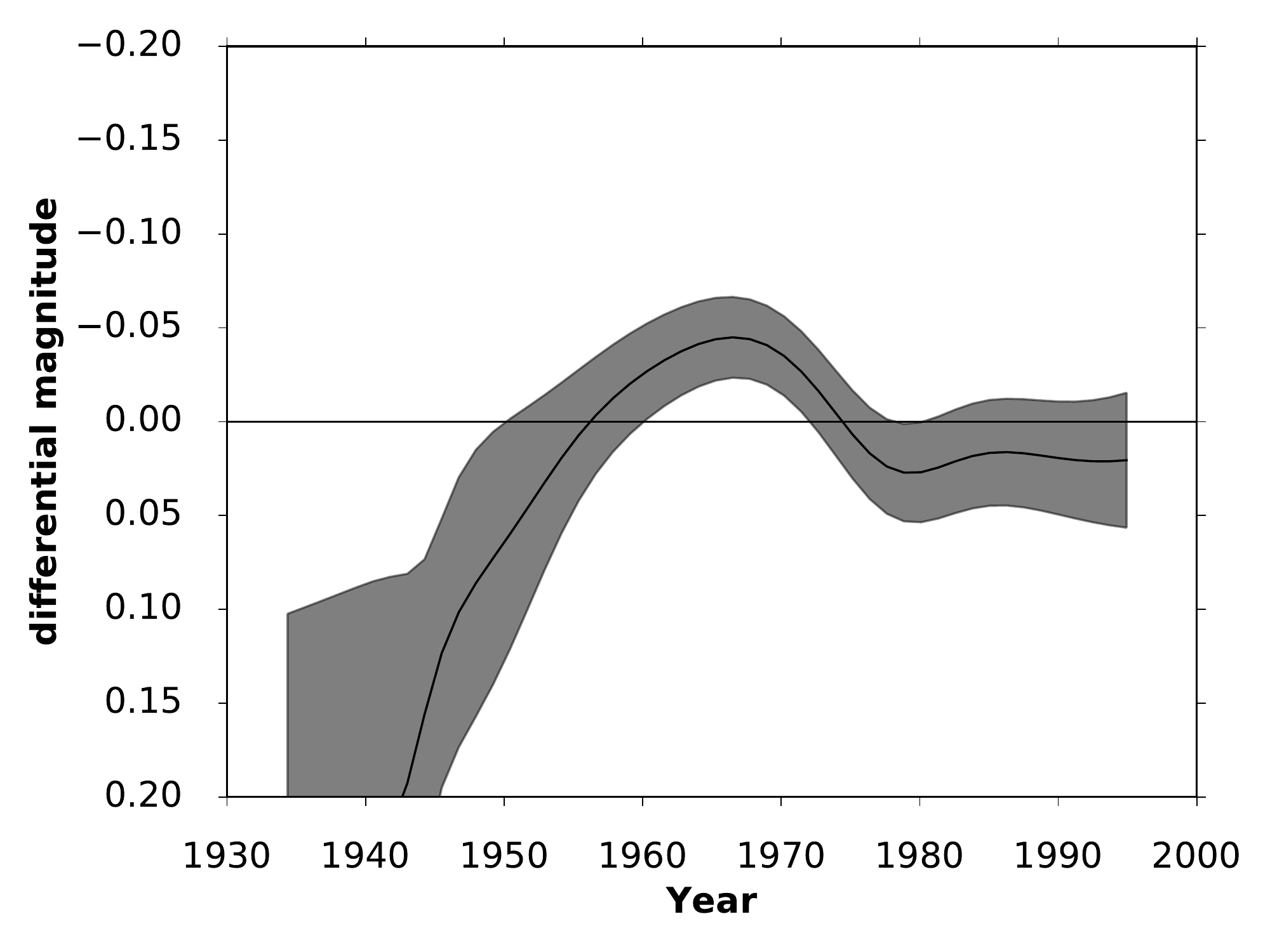}
\caption{\label{fig:kernel}Sonneberg data. Top, middle: Local polynomial smoothing using a Gaussian kernel function with a 5-year half-width (solid line) with $1\sigma$ uncertainties (shaded areas). The analysis was performed on the individual data points, but for visibility we show the 5-year bins only. Bottom: $pg-pv$ giving the color difference (redder color in positive values). Note that no $pv$ data are available for the 1940s; the trend shown at this time is fitted by the algorithm but can not be used for interpretation.}
\end{figure}

\begin{figure*}
\includegraphics[width=0.5\textwidth]{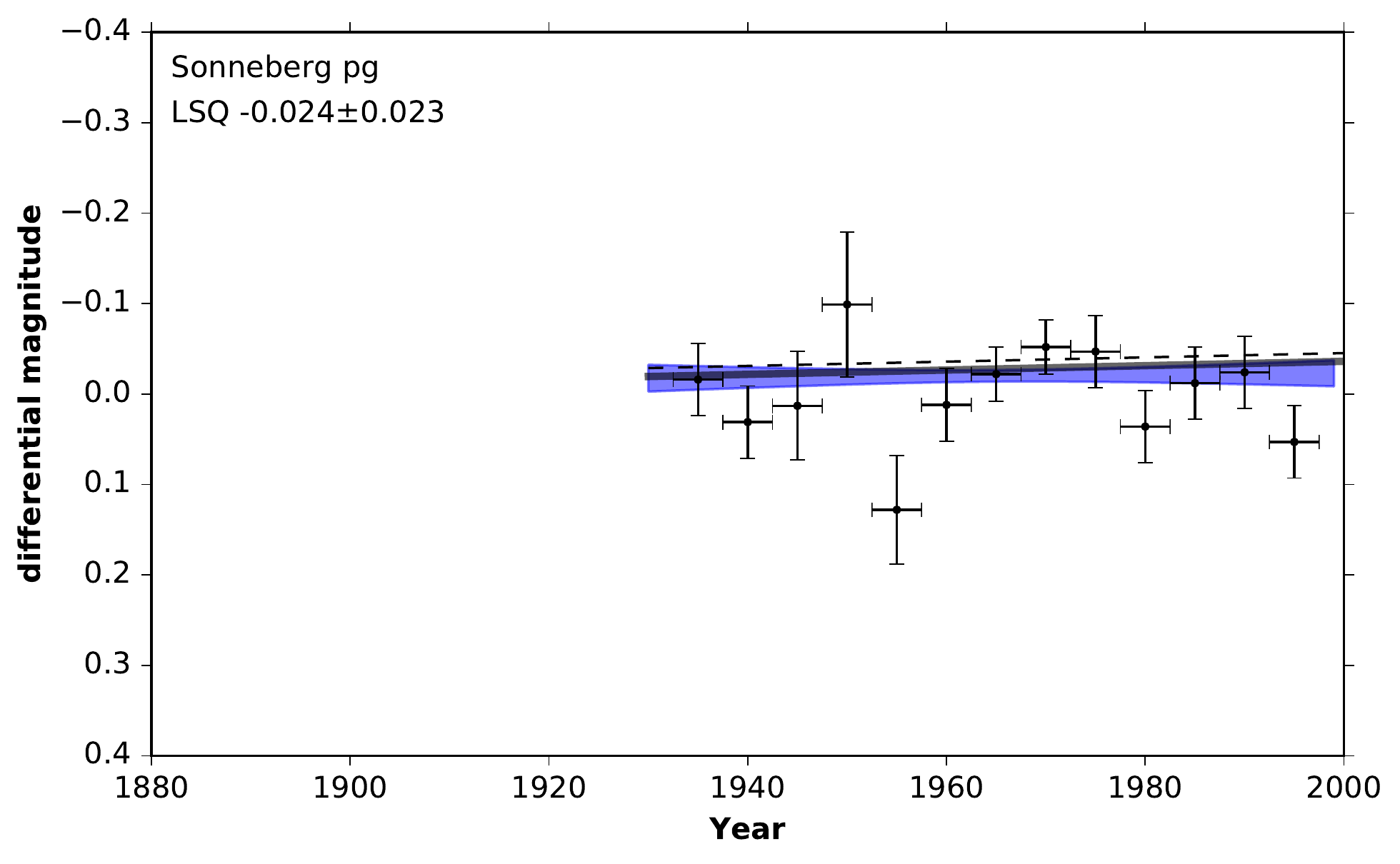}
\includegraphics[width=0.5\textwidth]{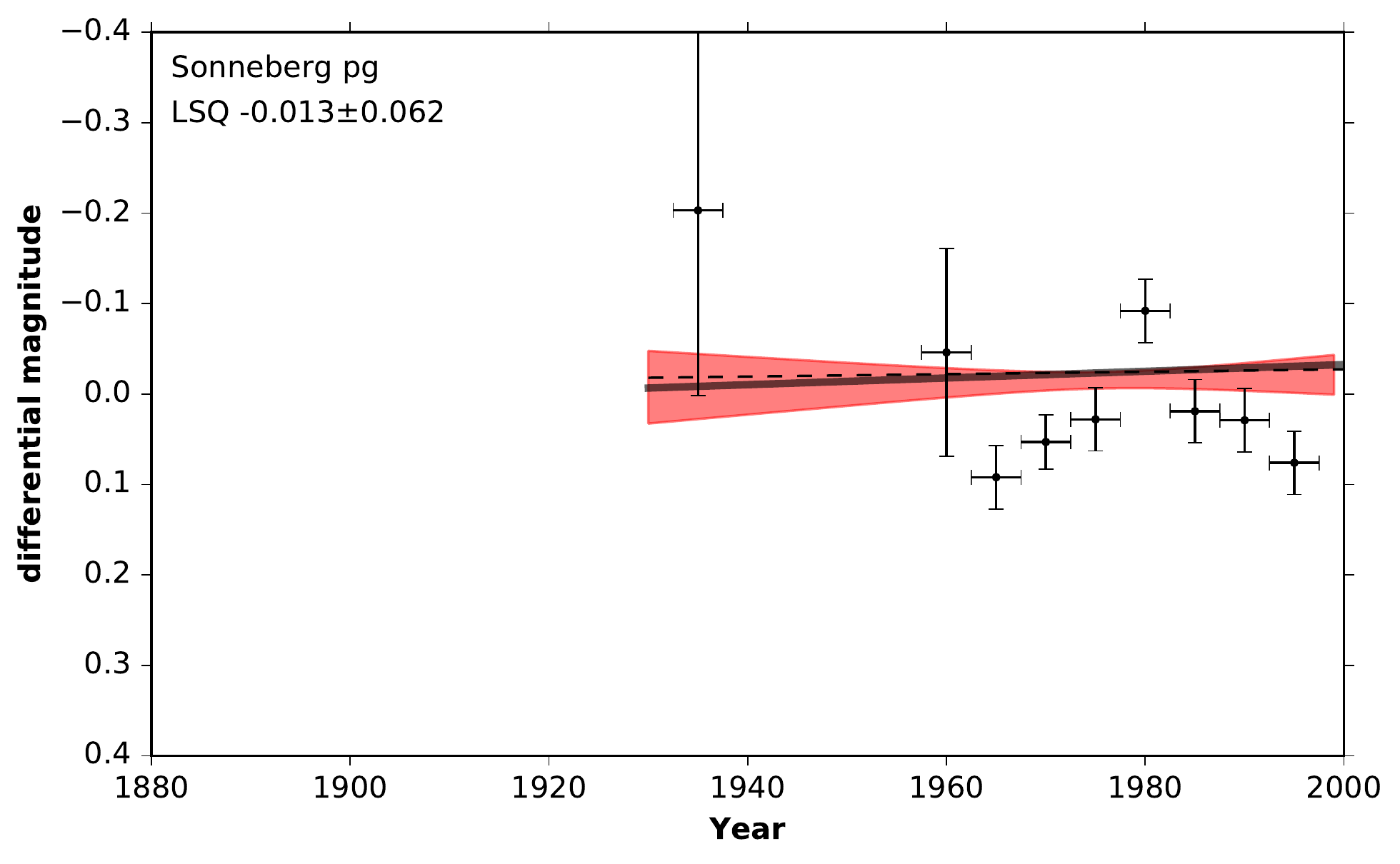}

\includegraphics[width=0.5\textwidth]{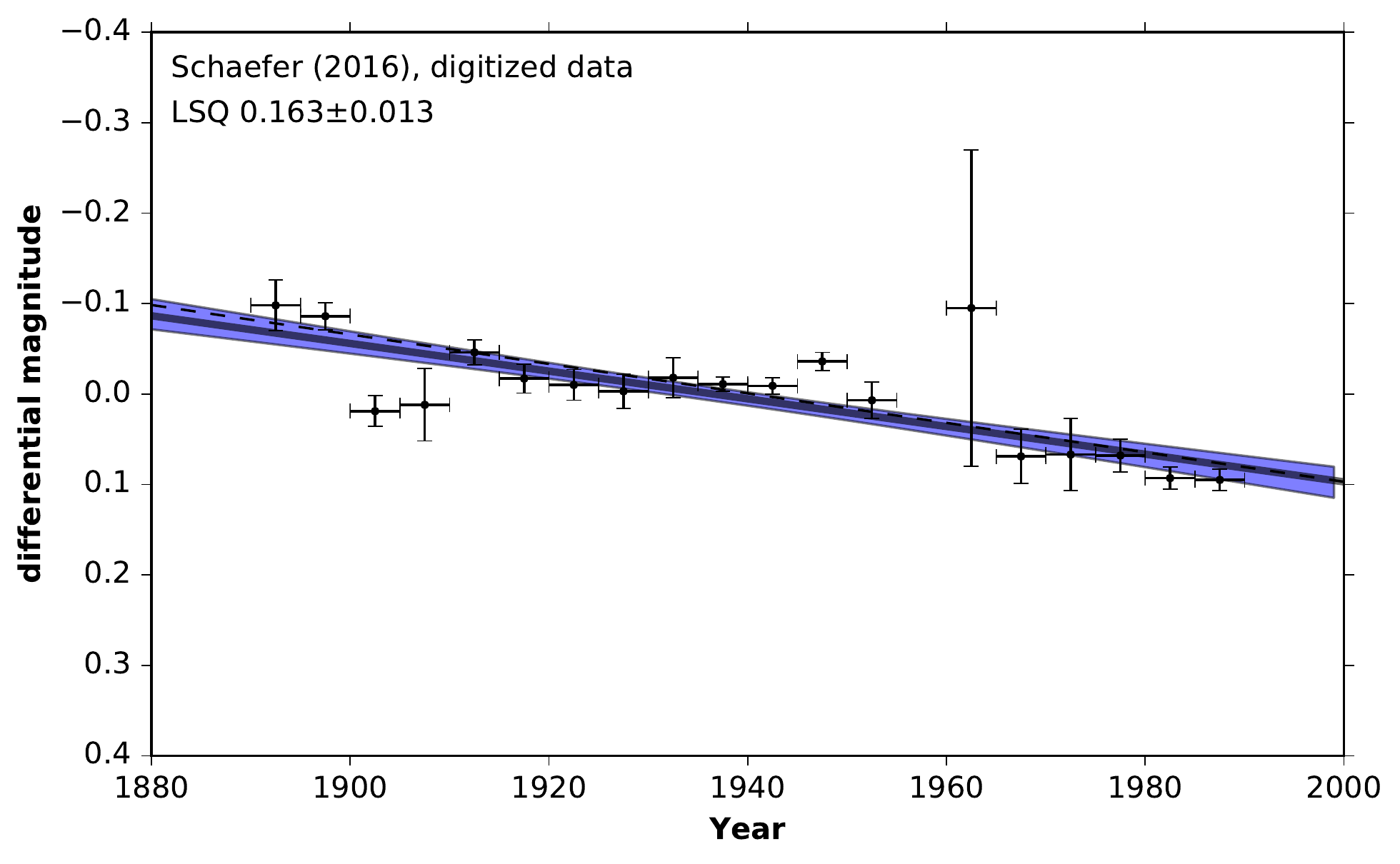}
\includegraphics[width=0.5\textwidth]{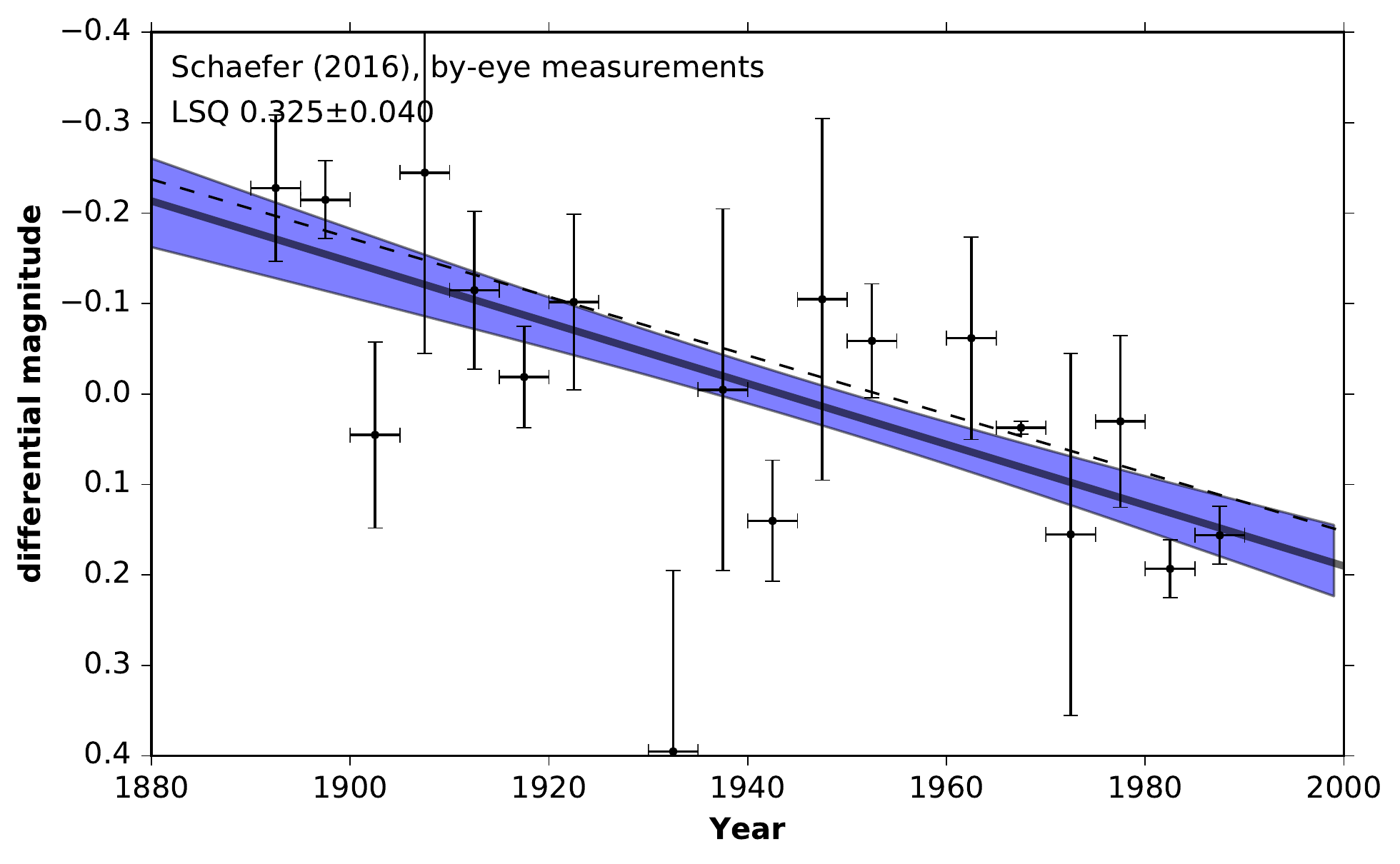}
\caption{\label{fig:comp}Comparison of Sonneberg and DASCH data using the same axes scales. Each plot shows the data in 5-year bins, a linear least-squares regression (dashed line), a robust regression using the maximum likelihood method (solid gray line), and a linear MCMC regression with $1\sigma$ uncertainties (shaded areas). The Sonneberg panels have been regressed with the individual data points; the 5-year bins are only shown for comparison. The data in the bottom panels are taken from Table 2 in \citet{2016ApJ...822L..34S}.}
\end{figure*}

\subsection{Check for seasonal variations and other meta-data}
Because of the location of the Sonneberg observatory ($50^\circ~22'~38.5''$N, $11^\circ~11'~23.3''$E), observation conditions are more favorable in the second half of the year. Indeed, the majority of plates (83\%) were taken between the months of June and November. During this time, the local climate changes from warmer summer temperatures ($>25^\circ$C) to colder winter temperatures ($<-5^\circ$C), while humidity is higher in autumn and lower in the summer and winter seasons.  We checked the magnitudes for a possible correlation with the observation month, perhaps caused by changes in temperature, humidity, or air mass. We sorted the months by their ave\-rage temperatures, and performed regressions versus the magnitudes. Also, we made monthly bins and compared their mean magnitudes to the overall mean. No significant correlation was found in either passband in any test. We conclude that the data do not show any obvious seasonal trend or variation. The same is true for other possible correlations such as the integration time.

\subsection{Trends on shorter timescales}
To check the data for trends on shorter timescales, we perform a local polynomial smoothing using a Gaussian kernel function with a 5-year half-width (Figure~\ref{fig:kernel}, top and middle panel). In parallel, we bin the data in 5-year bins for comparison with \citet{2016ApJ...822L..34S}.

Because of the non-Gaussian errors, the uncertainties of bins do not scale with $\sqrt{N}$. A Monte Carlo binning test reveals uncertainties of 0.05~mag in 5-year bins of $pv$ data, and 0.04~mag for $pg$. We use the larger uncertainties for the 5-year bins in Figure~\ref{fig:kernel}, and leave the local polynomial smoothing with Gaussian uncertainties.

For $pg$, we find a marginally significant ($\sim2\sigma$) brightening around 1970 in the smoothed polynomial. At this time, DASCH and $pv$ show nominal flux. This could be interpreted either as noise in the $pg$ data, or, if astrophysical, as a blue-shift color-change (Figure~\ref{fig:kernel}, bottom panel). As the polynomial underestimates the uncertainties, it is more likely to be noise.

The DASCH data suggest that the main dimming, if real, occurred between 1955 and 1965. This is not seen in $pg$ (and $pv$ has insufficient data for this time).

For the time between 1980 and 1995, we find a dimming trend in both colors (marginally significant at $\sim2\sigma$). However, as a similar brightening is seen in $pg$ for earlier times, which is not reflected in DASCH data, this is most likely because of remaining systematics in the data.

During other times, deviations from nominal flux are minor and can not be considered significant within our limits of $\sim0.05$~mag ($\sim5\%)$ per 5-year bin/half-width.

We conclude that we find no evidence for sub-trends within out limits of 5\% over 5-year segments.

\subsection{Literature comparison}
\label{sec:litcomp}
We test the hypothesis of a trend of 0.164~mag (DASCH, 0.325~mag by-eye) per century as claimed in \citet{2016ApJ...822L..34S} using an F-test for the $\beta$-parameter (slope) of the regression. The result indicates that a slope of 0.164~mag per century is in disagreement with the data by $4.9\sigma$, and $11\sigma$ for his by-eye estimate. This applies to $pg$, to $pv$, and to the combined data. The discrepancy with the claimed dimming is visually obvious as shown in Figure~\ref{fig:comp}.

\subsection{Light curve with Sternberg data}
The plate archive of the Sternberg Institute is the most important plate archive in Russia. It contains a total of 64,300 plates, of which the majority was taken with the 40cm Astrograph between 1950 and 1995 \citep{2006vopc.conf..103S}.

We searched the archives for all plates which contain Boyajian's star, and found a total of 156 plates taken with three different telescopes. Plates from the 40cm Astrograph ($n=17$) are the most accurate ones, with individual uncertainties of order 0.05~mag. The 16cm Tessar (series T) plates ($n=118$) suffer from systematics because of more than 10 different plate centers. This series contains the very last estimate from the Tessar equatorial camera at Moscow Presnya observatory. Finally, there are 21 plates taken with the 9.7cm Steinheil telescope, including the 4th-oldest plate taken from Moscow (12 Oct 1895). The uncertainties for the Tessar and Steinheil plates are 0.1--0.2~mag.

We note that there were three plates taken with the Astrograph on the evening of 24 Oct 1978 with mid-exposures at 18:00 UT, 18:40 UT, 19:15 UT. Our by-eye estimates are, in succession, 12.36, 12.33, and 12.28~mag, a brightening of 0.08~mag (8\% over 75 minutes). These plates are all of excellent quality, taken without moving the telescope between exposures. The emulsion was the same, one of the best ORWO emulsions of all times, and the comparison stars were very close to the variable. The program star appears clearly different in its brightness, and thus we consider this brightening to be significant and reliable (see appendix Figure~\ref{fig:sternscan}).

When compared to the 4 years of Kepler data, the fastest brightening observed (a dip egress) was 1.5\% per 30min cadence, or 3.5\% over 75 minutes. The occurrence rate of steep brightenings in Kepler data is low, $35$ of $65,316$ cadences show a brightening of $>0.5\%$. There are two nights with multiple observations from the Astrograph, so that the probability of observing such an event by chance is low (1 in 1000), if the occurrence rate of dips is constant over time. The deepest Kepler dips occurred with a separation of $\sim731$~days (2 years), and the last observed dip was on 1 March 2013. If the dips are periodic, it would require a period of 738 days to fall on 24 Oct 1978, with the next dip expected around 16 March 2017.

Regarding the light curve, with all Sternberg data we obtain an insignificant dimming of $0.046\pm0.040$~mag per century. If only the data from the Astrograph were used, we obtain $0.09\pm0.02$~mag per century, but the number of data points is low ($n=17$) and not well sampled (Figure~\ref{fig:obs}), covering only the time from 1960 to 1990. We believe that the slight trends are based on emulsion differences.

We conclude that the Sternberg result is consistent with no dimming of the star between 1895 and 1995 and inconsistent with the claimed dimming in \citet{2016ApJ...822L..34S}.

\subsection{Light curve with APPLAUSE data}
The APPLAUSE project digitizes photographic plates in the archives of Hamburg, Bamberg and Potsdam observatories \citep{2014aspl.conf...53G,2014aspl.conf..127T}. The plates are digitized with high-resolution flatbed scanners. As of August 2016, 42,789 plates have been scanned and reduced\footnote{\hyperref[]{http://dx.doi.org/10.17876/plate/dr.2/},\\accessed 26-August 2016.}. The data have been used e.g. in a study on the historic long-term variability of active galaxies \citep{2016arXiv160700312W}. We have extracted the magnitudes for Boyajian's Star using the public web interface. A first analysis showed a correlation of the star's brightness with its position on the plate, which will be corrected in the next data release. For Boyajian's Star, this accounts for 0.15~mag extra scatter. We have reduced the scatter by subtracting the mean magnitude of the annular bin (there are 8 annular bins). For a least-squares linear regression, this is irrelevant, as it does not change the slope of the fit. The result is a dimming of $0.61\pm0.21$~mag/century, inconsistent with all other data from Harvard, Sonneberg, Sternberg and Pulkovo (Figure~\ref{fig:obs}).

The slope in the APPLAUSE light curve is strongly affected by the data points from 1938--1939, which originate from a telescope at the Hamburg Observatory that has a strong coma effect in the images. The other data points come from three other telescopes. The APPLAUSE data are heterogeneous, with various telescopes and a wide range of emulsions and exposure times used, so calibration is an issue. Also, photo\-metry has been calibrated globally, without choosing local comparison stars for Boyajian's Star. Such calibrations might be performed in a future data release. At the moment, however, the APPLAUSE data suffer from large systematics in the case of Boyajian's Star and can not be used for an individual analysis.

\subsection{Light curve with Pulkovo data}
Pulkovo observatory is the main astronomical observatory of the Russian Academy of Sciences, located outside of St.~Petersburg. It was founded in 1839 and the regular photographic observations were started at in 1894. The collection today contains 50,000 plates \citep{2002rict.confE..22K}. The plates were measured using the ``Ascorecord'' machine with visual pointing, the ``Fantasy'' automatic complex \citep{1989WisZe..38...28G}, and the UMAX scanner (600 ppi) \citep{2016AstL...42...41I}. In 2010, the measurement and calibration technique for wide fields digitized by the scanner Microtec (3200 ppi) was developed. About 2000 plates with asteroids observations were digitized and measured \citep{2011gfun.conf..131K}. The data were used for e.g. studies on Pluto \citep{2013arXiv1310.7502K}. The plates were taken with the ``Normal Astrograph'', with 330 mm aperture and 3467 mm focal length, between 1922 and 2001. The plates' emulsion had a maximum sensitivity close to the B-band (440-450 nm). For Boyajian's Star, we decided to use a digital camera for plate digitization. The plate digitization and calibration process is described in \citep{2016AstL...42...41I}  and used 9 nearby stars for calibration. The ave\-rage standard deviation from the mean is 0.13~mag. A least-squares regression yields $0.07\pm0.22$~mag/century (Figure~\ref{fig:obs}, bottom). Because of the low number of data points, any one point already changes the result significantly. Therefore, the Pulkovo data can not be used for an individual analysis of Boyajian's Star.

\subsection{Light curve with DASCH data}
The data were discussed extensively in \citet{2016ApJ...825...73H}. For comparison, we show all APASS-calibrated data without quality cuts, resulting in a slope of 0.12~mag instead of 0.16~mag per century as suggested by \citet{2016ApJ...822L..34S} (Figure~\ref{fig:obs}, top).

\subsection{Period search}
\label{sec:periods}
\citet{2016MNRAS.457.3988B} discuss several periodic signals (their section 2.1). The rotation period of the star is found to be 0.88~d \citep{2016MNRAS.457.3988B}, although this period has been claimed to originate from a different star \citep{2016ApJ...833...78M}. From the separation of dips, a possible period of a circumstellar body is identified, at 48.4 or 48.8~d (or half this value). Finally, the largest dips are separated by 726~d. In order to search for such periodicities in our historic data, we combine all photo\-metry (DASCH, Sonneberg, APPLAUSE, Pulkovo), and normalize the flux values per observatory by their median. Each measurement (plate) received the same weight.

For sinusoidal trends, such as the rotation signal, the Lomb-Scargle (LS) periodogram is most sensitive \citep{1982ApJ...263..835S,1986ApJ...302..757H,2009A&A...496..577Z}. For transit-like events, the Box-fitting Least Squares (BLS) algorithm \citep{2002A&A...391..369K} is more suitable. The BLS has been run for periods up to 50 years, and transit durations between 1 hour and 10 days.

Both methods, LS and BLS, find only peaks that are yearly aliases, and a few caused by single outliers. When they are ignored, no significant peaks remain. We test the sensitivity to a 48.8~d transit signal through injection and retrieval. A 10\% dip of 1-day length, repeating regularly every 48.8~d, would have been detected with $>3\sigma$ confidence. Consequently, we can constraint transits to $<10\%$ in depth for periods $<50$~d. For 726-day signals, we can not give useful limits because of insufficient data caused by many gaps.

\section{Discussion}
\label{sec:discussion}
The claim of a century-long dimming by \citet{2016ApJ...822L..34S} is weakened by the fact that only 131 eye estimates were made for KIC~8462852 from the Harvard plates, which represents merely a few hours of plate scanning by an experienced observer. It is not unusual to perform 500 or more estimates on one day. To substantiate such a strong claim from by-eye measurements, we would recommend estimating all available plates in a controlled test (blind and shuffled). Another weakness is the use of DASCH estimates, which were calibrated globally by stars located all over the plates, many of which lie well away from the object of interest. Patrol series plates, for example, measure 50 degrees by 40 degrees, and DASCH measures from them can differ systematically by $\pm0.3$~mag.

If we take the results of constant flux (within the limits of $\sim5\%$ per 5-year bin) at face value, it is still possible that the star undergoes periodic changes in brightness. A brightness variation of $\sim3\%$ over a few years, as reported by \citet{2016ApJ...830L..39M}, is compatible with the Sonneberg (and DASCH) data, and invisible in their noise.

\citet{2016arXiv161003219L} recently discussed possible time spans for a potential dimming phenomenon. For example, if the star would have dimmed continuously as it likely did during the Kepler mission time, it would have been a million times brighter (V=2.45) in ancient times (2,500 years ago), and it should appear in star catalogs of this time (which we checked, and which is not the case). Also, such a brightness variation is in disagreement with the star being of type F3V in a distance of $391.4\substack{+53.6 \\ -42.0}$~pc \citep{2016arXiv160905492H}. With the Gaia distance estimate, and the known luminosity of an F3V star, we can constraint the brightness to be nominal within $\sim20\%$.

Consequently, the duration of brightness changes might be larger than a few years, but shorter than 2,500 years (and shorter than $\sim150$ years, if the Gaia result is not affected from systematics). Perhaps the star exhibits periodic decade-long fluctuations at a level of a few percent.

\section{Conclusion}
We have analyzed photo\-metry from the Sonneberg observatory (1934 -- 1995), and other sources (1895 -- 1995), and found strong evidence that Boyajian's Star has not dimmed, but kept a constant flux within a few percent. Variations on 5-year scale are found to be of order 5\% or less. We found a possible $>=8\%$ dip on 24 Oct 1978, resulting in a period of the putative occulter of 738 days.

\acknowledgments
T.T. acknowledges support by the Estonian Ministry of Education and Research (IUT26-2, IUT40-2) and by the European Regional Development Fund (TK133). N.S. and E.P. acknowledge support from the P-7 basic research program of the Presidium of Russian Academy of Sciences. I.B. acknowledges plate expert assistance by Radislav Dorozkin and Vyacheslav Levitsky.

\software{Astrometry.net \citep{2010AJ....139.1782L}, SExtractor program \citep{2006ASPC..351..112B}, emcee toolkit \citep{2013PASP..125..306F}.}

\appendix

\begin{figure*}
\includegraphics[width=0.5\textwidth]{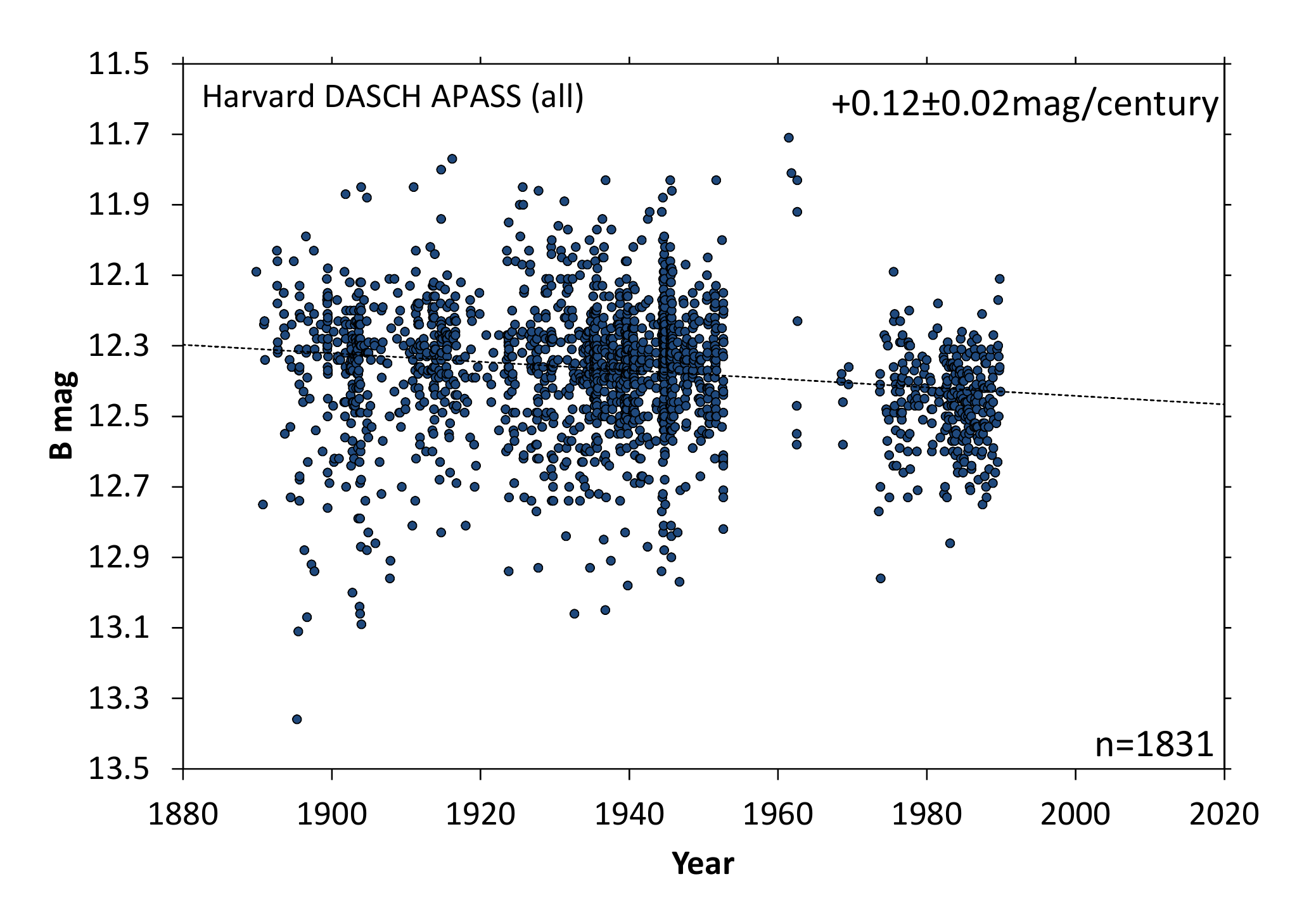}
\includegraphics[width=0.5\textwidth]{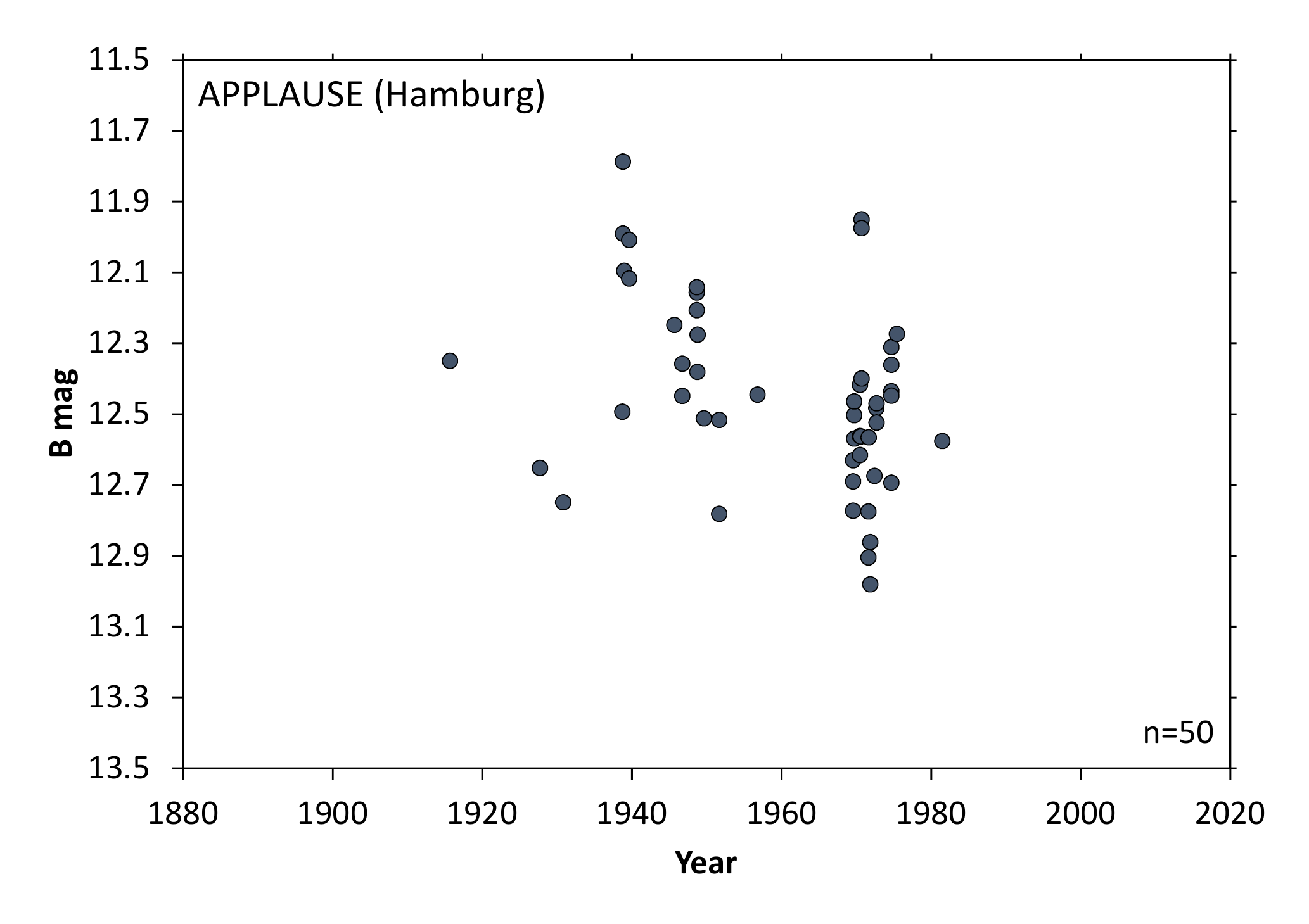}

\includegraphics[width=0.5\textwidth]{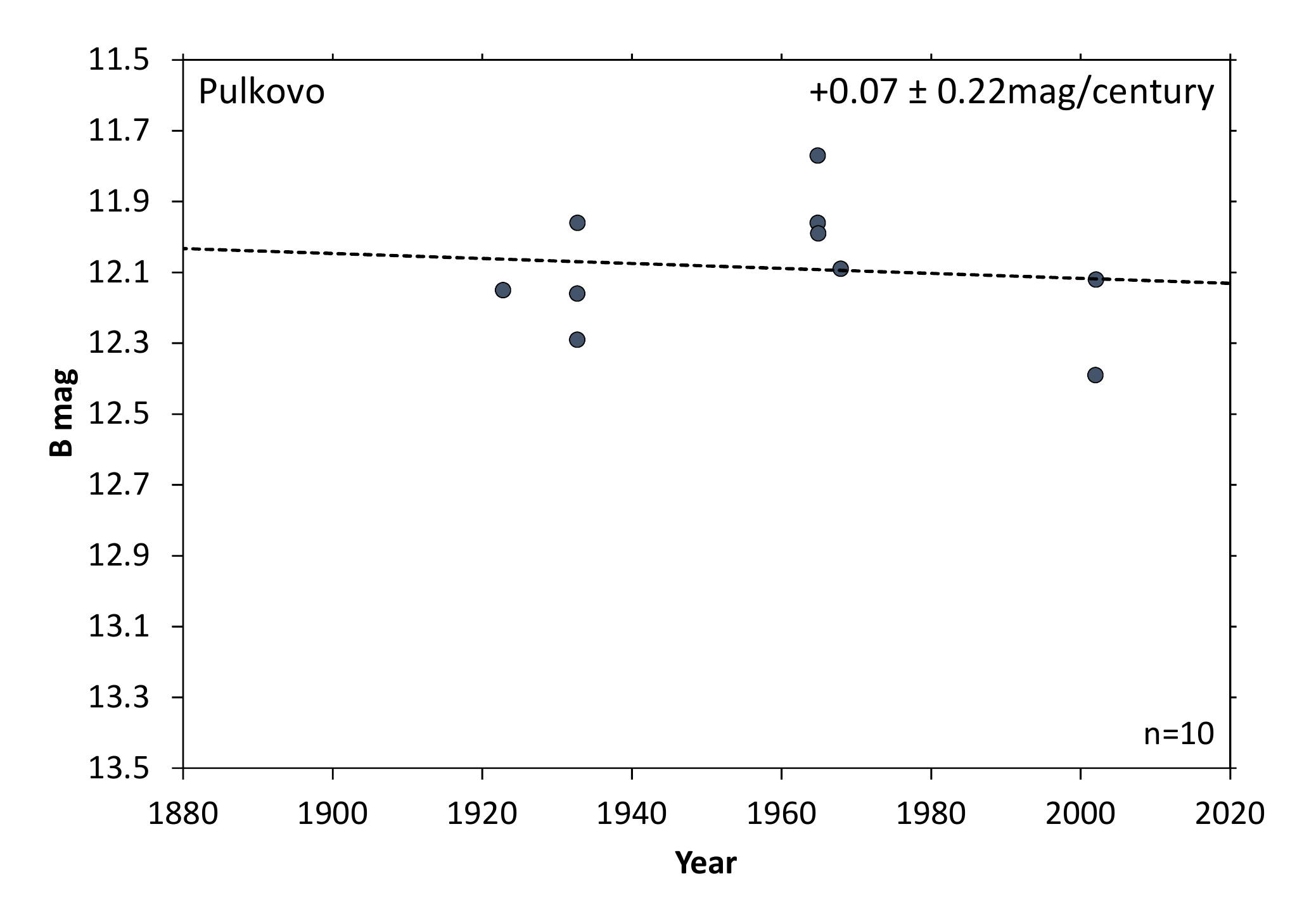}
\includegraphics[height=0.3\textwidth]{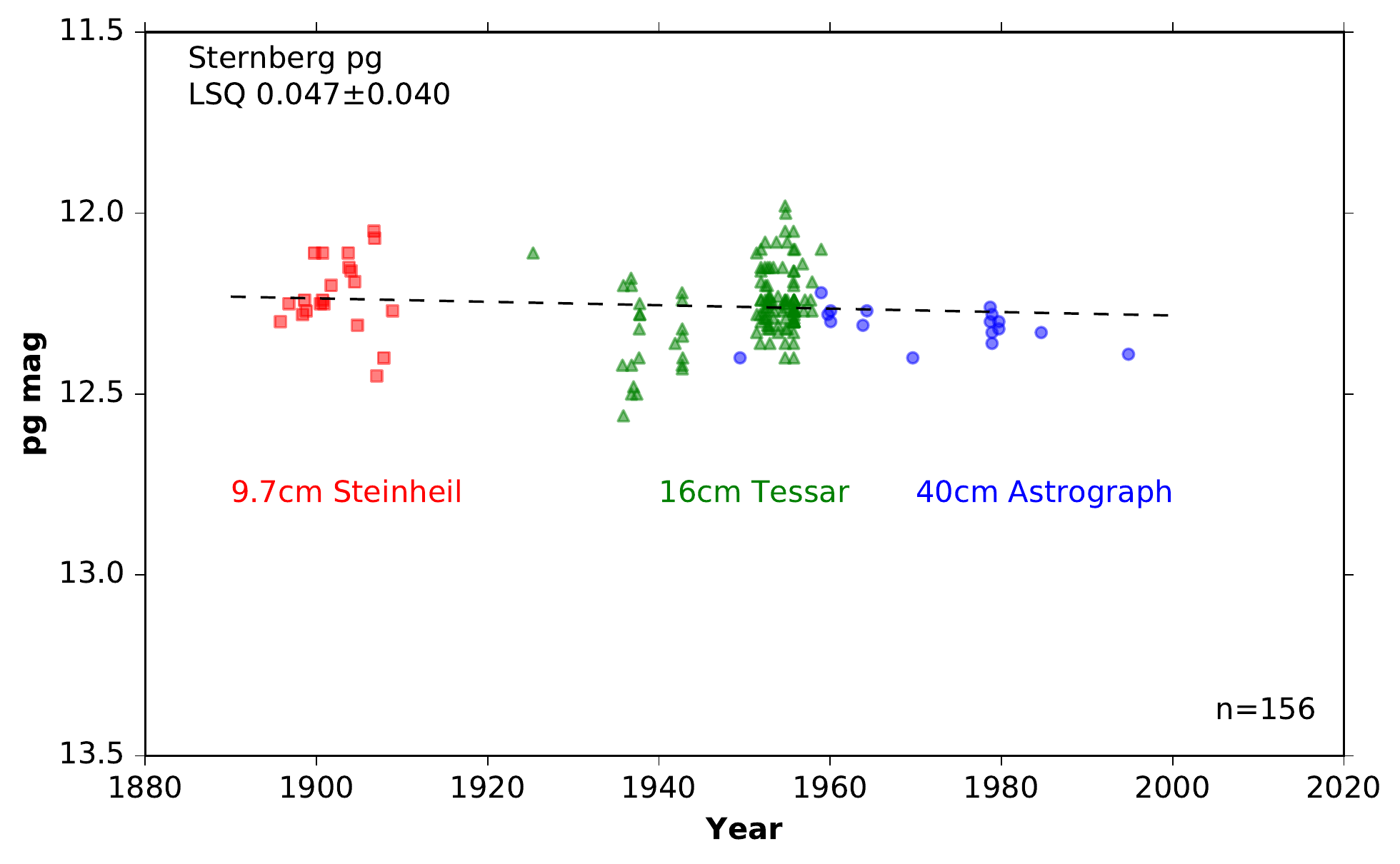}
\caption{\label{fig:obs}Photo\-metry from DASCH (top left), APPLAUSE (top right), Pulkovo (bottom left) and Sternberg (bottom right).}
\end{figure*}

\label{sec:appendix}
\begin{figure*}
\includegraphics[width=0.5\textwidth]{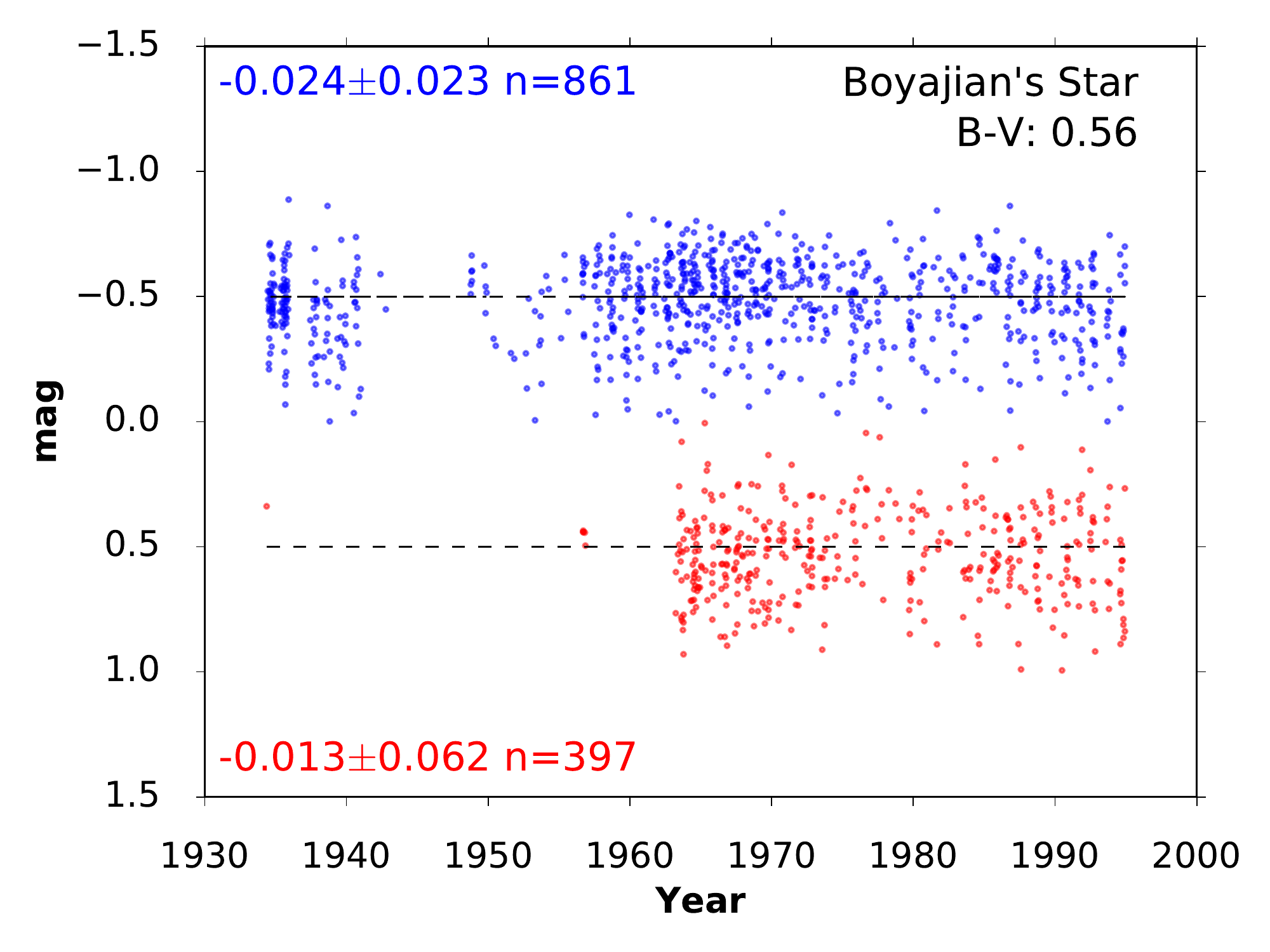}
\includegraphics[width=0.5\textwidth]{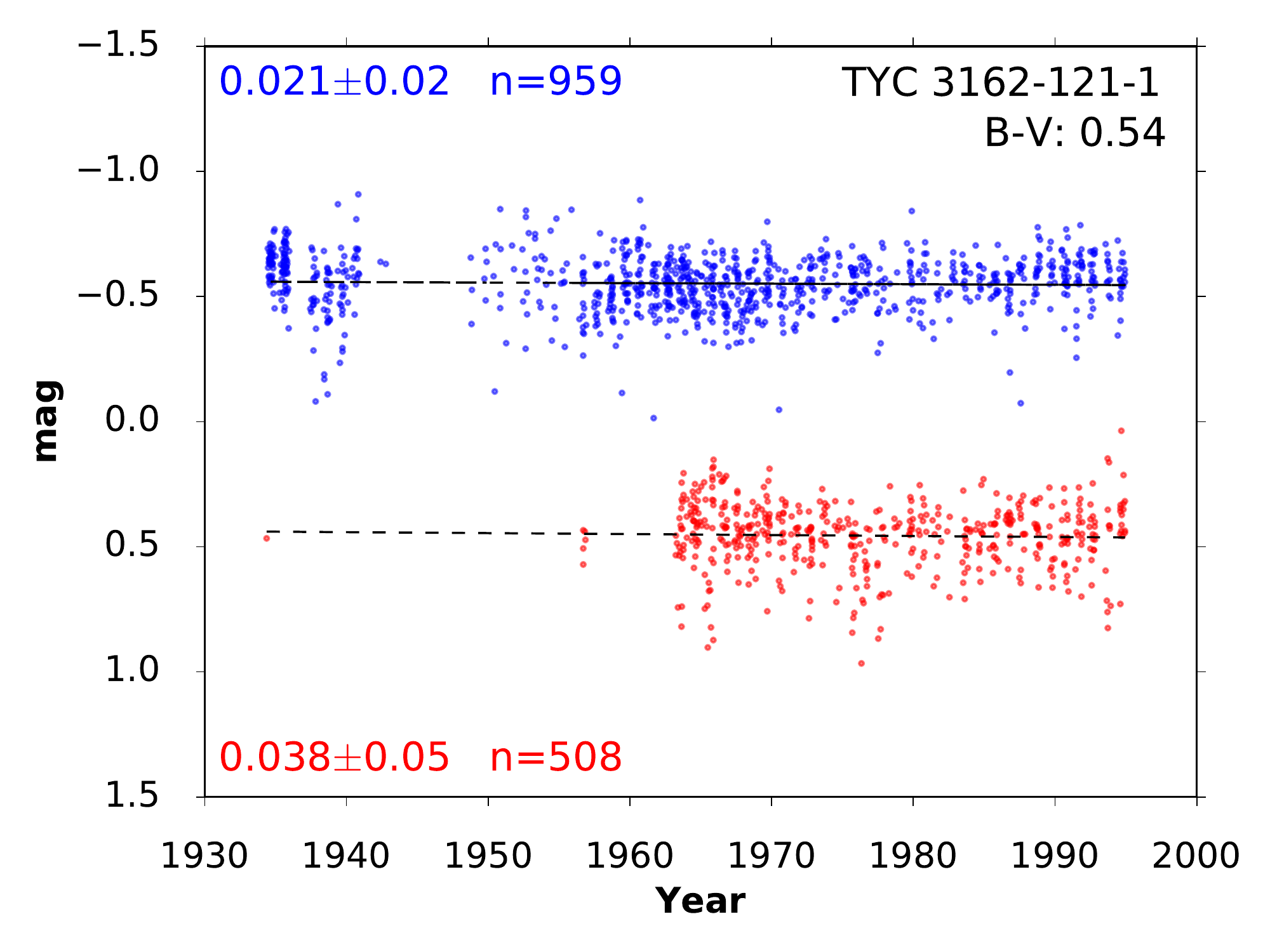}

\includegraphics[width=0.5\textwidth]{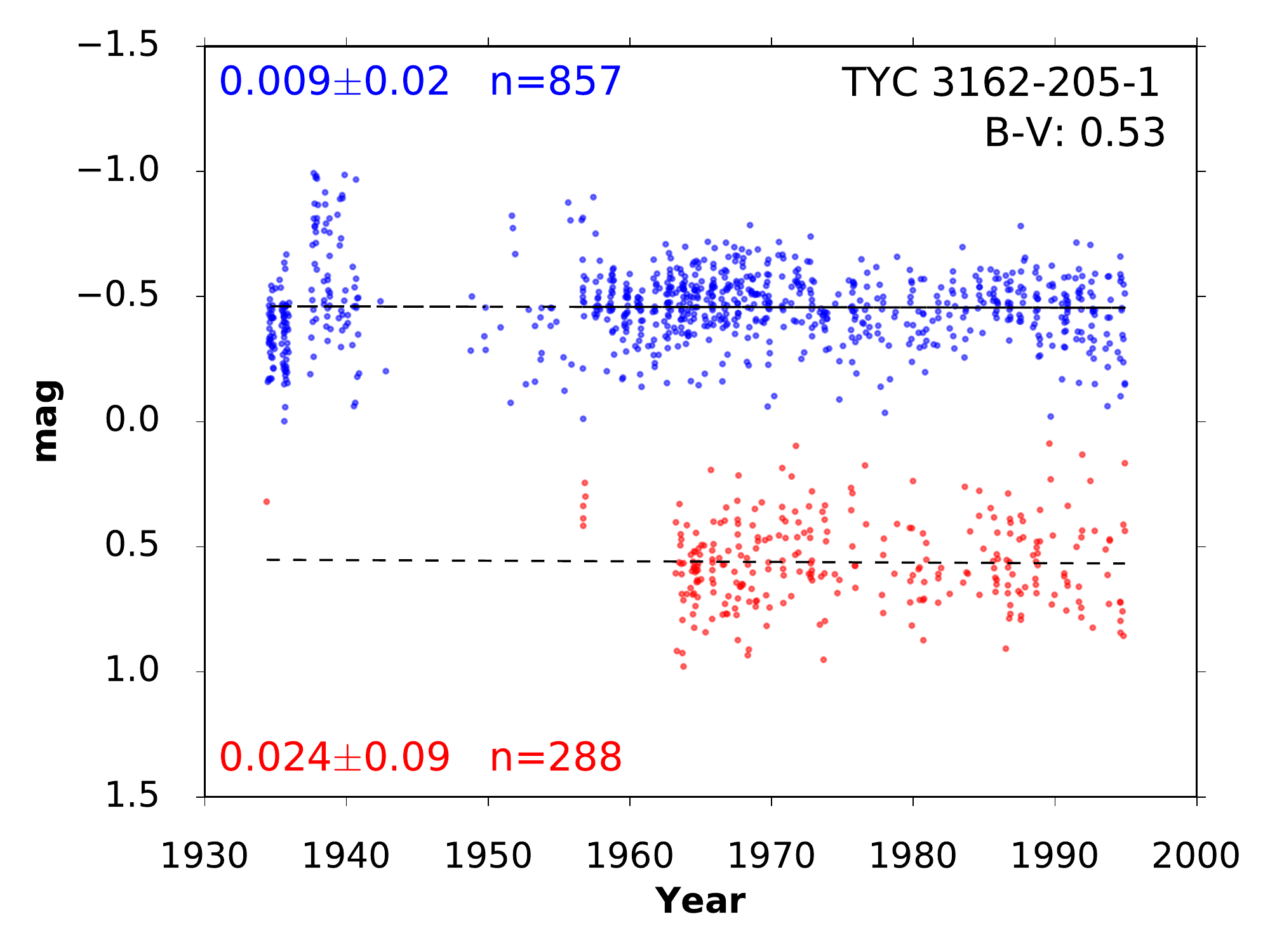}
\includegraphics[width=0.5\textwidth]{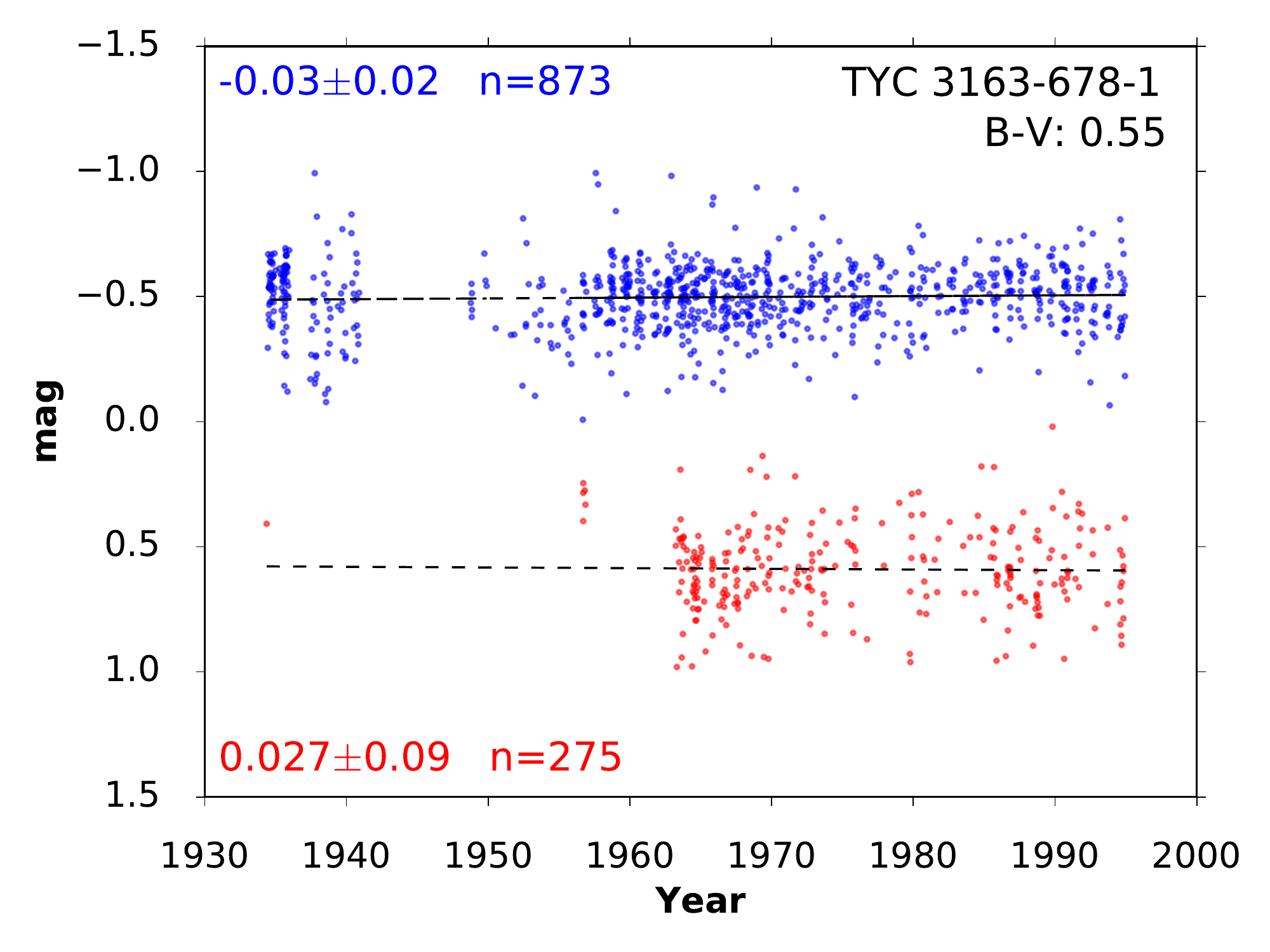}

\includegraphics[width=0.5\textwidth]{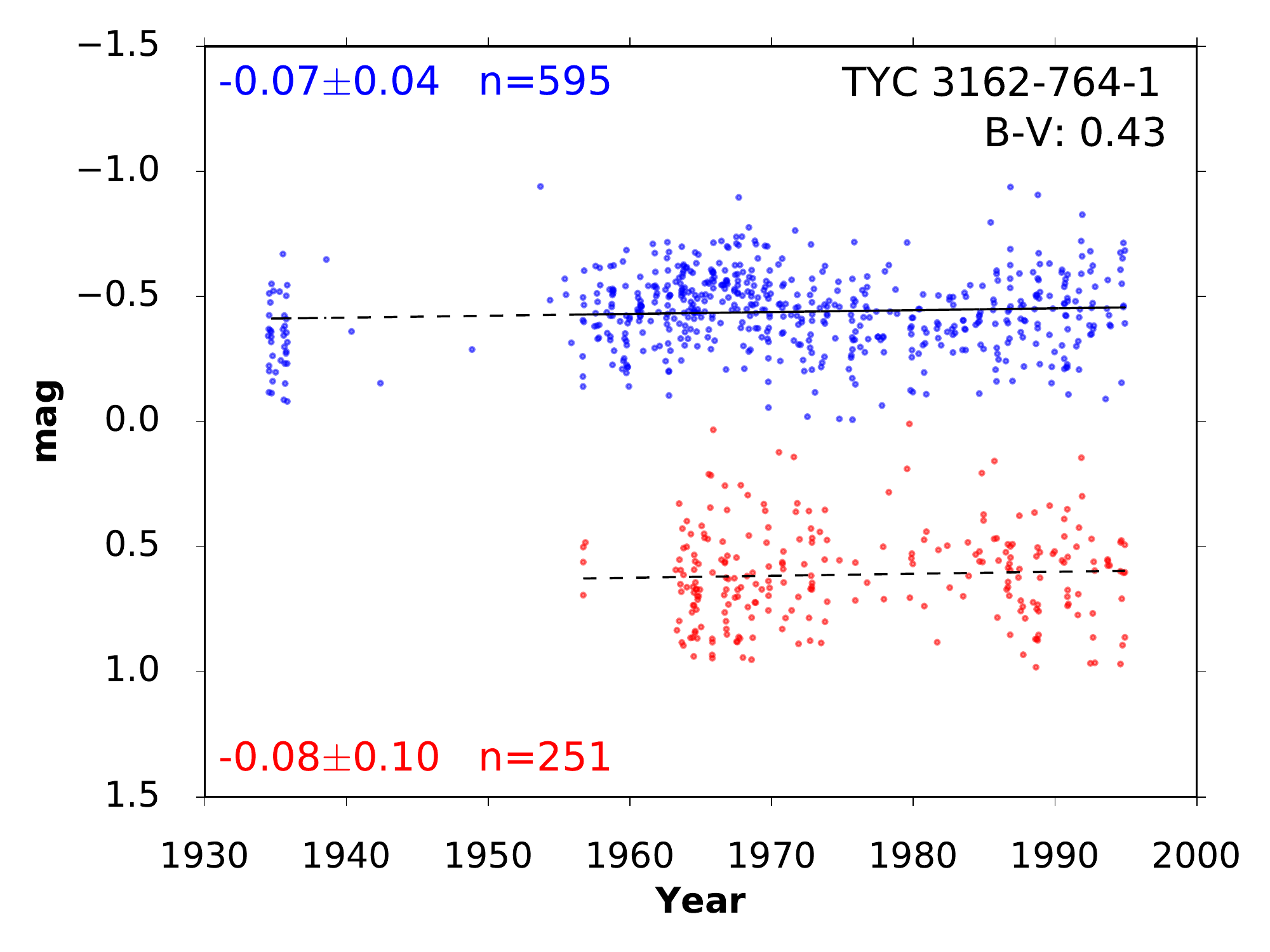}
\includegraphics[width=0.5\textwidth]{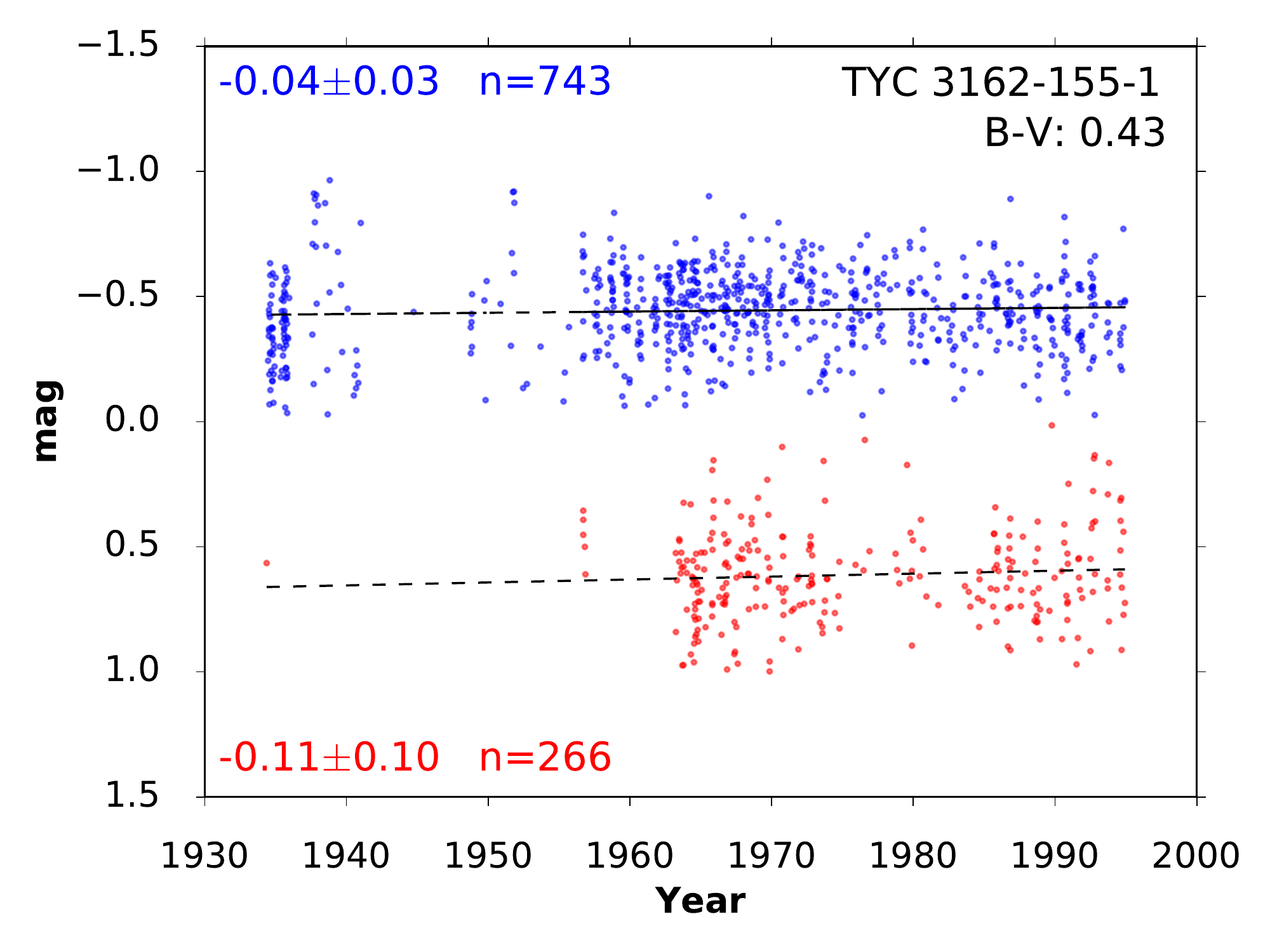}
\caption{\label{fig:appendix}Sonneberg examples of comparison stars. The program star, Boyajian's Star, is shown on the upper left for comparison. Differential magnitudes have been offset to -0.5 for $pg$ and +0.5 for $pv$ for better visibility.}
\end{figure*}

\begin{figure*}
\includegraphics[width=0.5\textwidth]{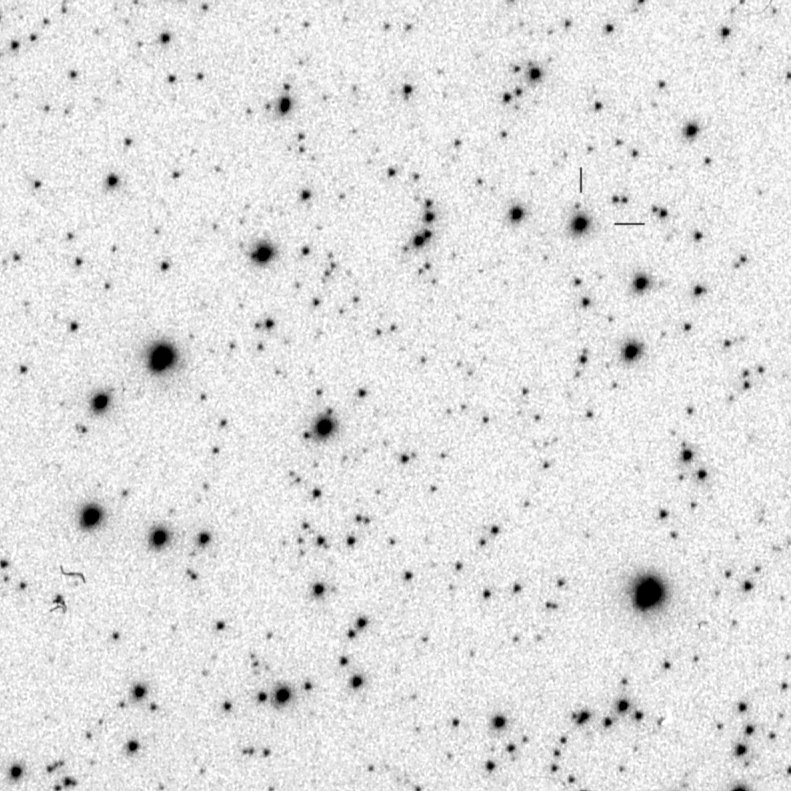}
\includegraphics[width=0.5\textwidth]{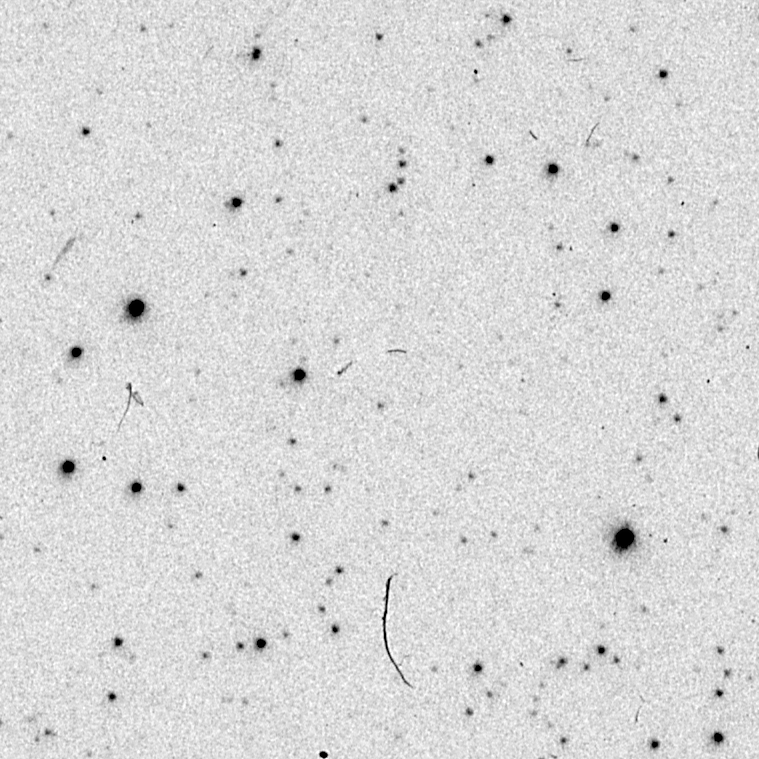}

\includegraphics[width=0.5\textwidth]{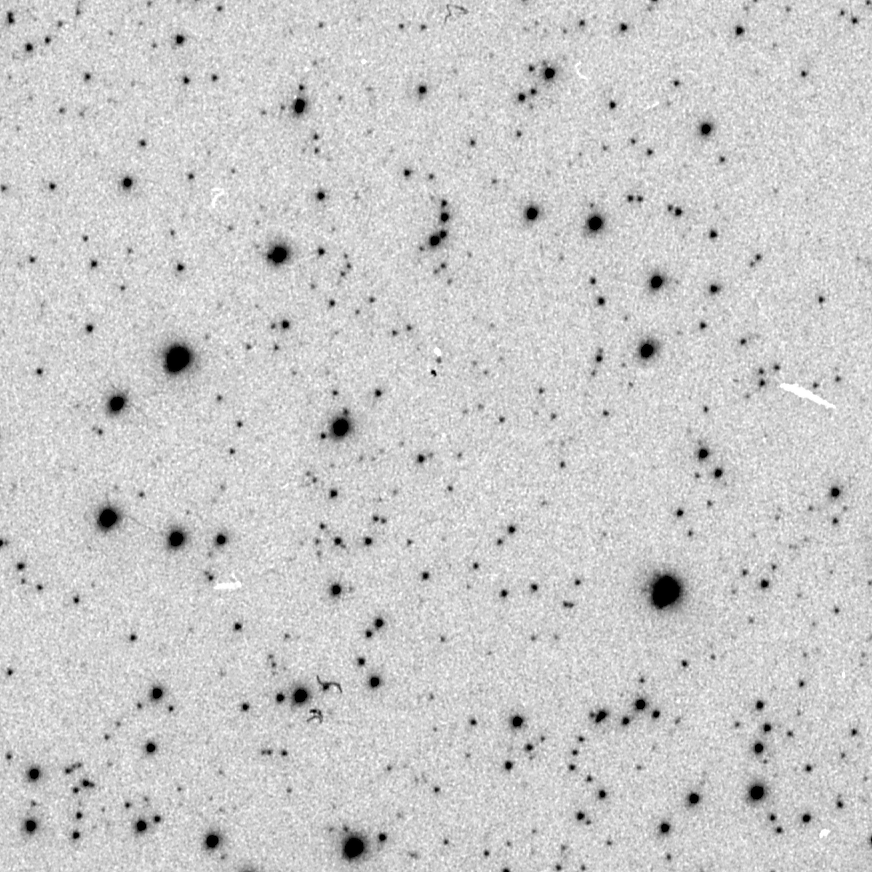}
\includegraphics[width=0.5\textwidth]{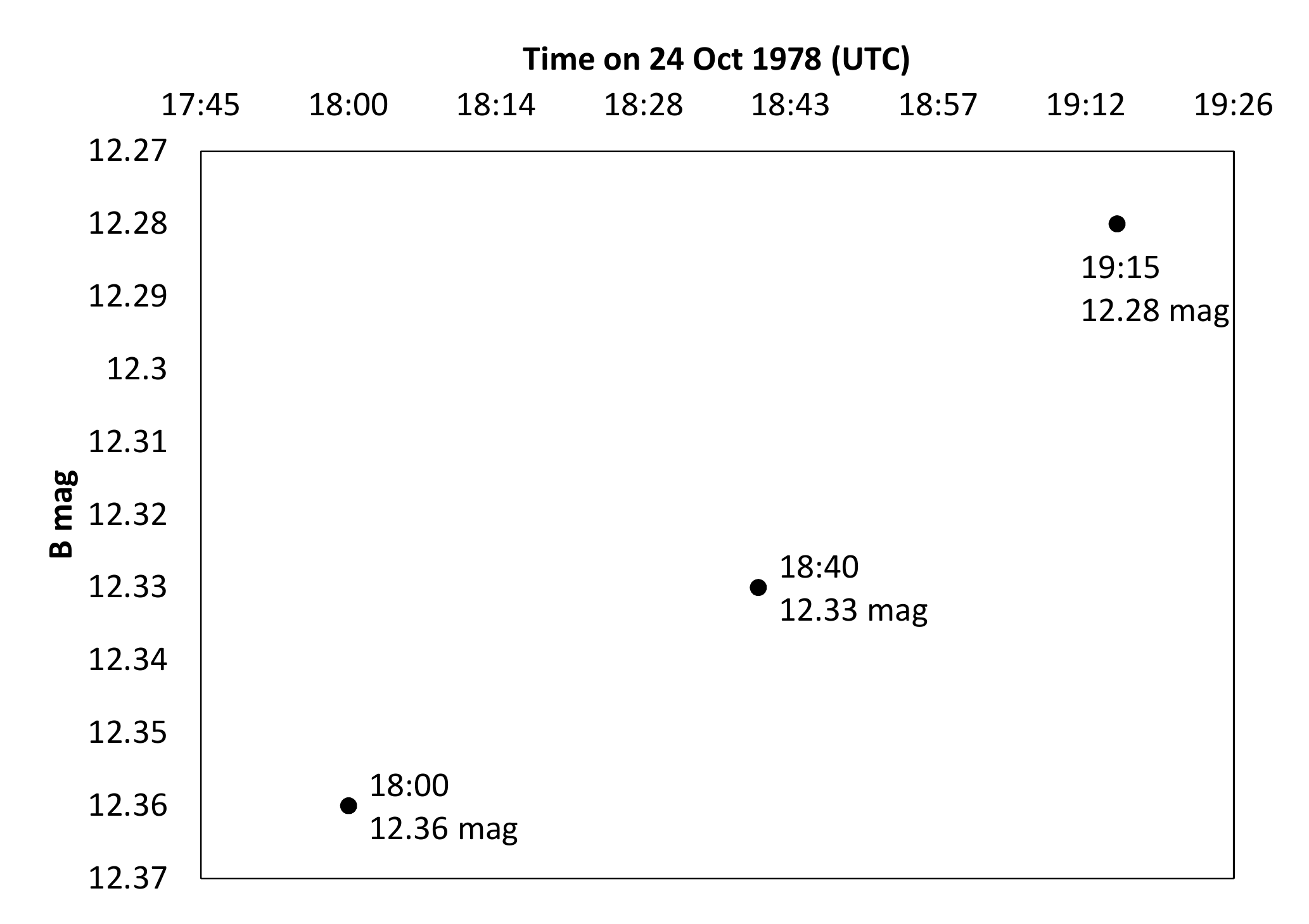}
\caption{\label{fig:sternscan}Sternberg Astrograph scans from 24 Oct 1978, taken at  18:00 UT (upper left), 18:40 UT (upper right), 19:15 UT (lower left). Our by-eye estimates are, in succession, 12.36, 12.33, and 12.28~mag.}
\end{figure*}

\pagebreak
\clearpage

\bibliography{sonneberg}
\bibliographystyle{yahapj}

\end{document}